\documentclass[conference]{acmart}
\IEEEoverridecommandlockouts

\usepackage{fancybox,framed}
\usepackage{fancyvrb}
\usepackage{balance}
\usepackage{makecell}
\usepackage{mathtools}
\usepackage{enumitem}
\usepackage{enumitem,kantlipsum}
\usepackage{fancyvrb}
\usepackage{listings}
\usepackage{xcolor}
\usepackage{subfigure}
\usepackage[ruled,linesnumbered,noend]{algorithm2e}
\usepackage{pgfplots}
\usepackage{adjustbox}
\usepackage{booktabs}
\usepackage{caption}
\usepackage{graphicx}
\usepackage{amsmath}
\usepackage{soul}
\usepackage{diagbox}
\usepackage{makecell}
\usepackage{multirow}
\usepackage{simplebnf}
\usepackage{amsthm}
\newtheorem{theorem}{Theorem}

\usepackage{colortbl}
\usepackage{url}
\usepackage{footnote}
\usepackage{tikz}
\makesavenoteenv{table}
\usepackage{pgfplotstable}
\usepgfplotslibrary{groupplots}

\def\BibTeX{{\rm B\kern-.05em{\sc i\kern-.025em b}\kern-.08em
    T\kern-.1667em\lower.7ex\hbox{E}\kern-.125emX}}
    
\definecolor{codegreen}{rgb}{0,0.6,0}
\definecolor{codegray}{rgb}{0.5,0.5,0.5}
\definecolor{codepurple}{rgb}{0.58,0,0.82}
\definecolor{backcolour}{rgb}{0.95,0.95,0.92}

\lstdefinestyle{mystyle}{
    backgroundcolor=\color{backcolour},   
    commentstyle=\color{codegreen},
    keywordstyle=\color{magenta},
    numberstyle=\tiny\color{codegray},
    stringstyle=\color{codepurple},
    basicstyle=\ttfamily\footnotesize,
    breakatwhitespace=false,         
    breaklines=true,                 
    captionpos=b,                    
    keepspaces=true,                 
    numbers=left,                    
    numbersep=5pt,                  
    showspaces=false,                
    showstringspaces=false,
    showtabs=false,                  
    tabsize=2
}

\begin{document}

\newcommand{\ziawasch}[1]{\textcolor{purple}{Ziawasch: #1}}
\newcommand{\za}[1]{\textcolor{purple}{ZA: #1}}
\newcommand{\ma}[1]{\textcolor{cyan}{Mahdi: #1}}
\newcommand{\schnell}[1]{\textcolor{orange}{Christoph: #1}}
\newcommand{\maedit}[1]{\textcolor{blue}{#1}}
\newcommand{\jq}[1]{\textcolor{olive}{JQ: #1}}
\newcommand{\rjm}[1]{\textcolor{green}{RJM: #1}}

\newcommand{\system}{\textsc{Blend}\xspace}
\newcommand{\tightlist}{\itemsep=-2pt}
\newcommand*{\rom}[1]{\expandafter\@slowromancap\romannumeral #1@}

\newcommand{\Cross}{\mathbin{\tikz [x=2.5ex,y=2.5ex,line width=.5ex] \draw (0,0) -- (1,1) (0,1) -- (1,0);}}

\newcommand{\answerRone}[1]{{\color{red} {#1}}}
\newcommand{\answerRtwo}[1]{{\color{blue} {#1}}}
\newcommand{\answerRthree}[1]{{\color{purple} {#1}}}
\newcommand{\answer}[1]{{\color{orange} {#1}}}

\renewcommand\theadfont{}
\DeclarePairedDelimiter\ceil{\lceil}{\rceil}
\DeclarePairedDelimiter\floor{\lfloor}{\rfloor}

\newcounter{enum}
\newenvironment{packed_enum}{
\begin{list}{\textbf{(\arabic{enum})}}{
  \setlength{\itemsep}{0pt}
  \setlength{\parskip}{0pt}
  \setlength{\labelwidth}{-5 pt}
  \setlength{\leftmargin}{0 pt}
  \setlength{\itemindent}{0pt}
  \usecounter{enum}}
}{\end{list}}

\title{BLEND: A Unified Data Discovery System}

\author{\IEEEauthorblockN{1\textsuperscript{st} Mahdi Esmailoghli}
\IEEEauthorblockA{
\textit{TU Berlin \& BIFOLD}\\
Berlin, Germany \\
esmailoghli@tu-berlin.de}
\and
\IEEEauthorblockN{2\textsuperscript{nd} Christoph Schnell}
\IEEEauthorblockA{
\textit{Leibniz Universität Hannover}\\
Hannover, Germany \\
schnell@dbs.uni-hannover.de}
\and
\IEEEauthorblockN{3\textsuperscript{rd} Ren\'ee J. Miller}
\IEEEauthorblockA{
\textit{University of Waterloo}\\
Waterloo, Canada \\
rjmiller@uwaterloo.ca}
\and
\IEEEauthorblockN{4\textsuperscript{th} Ziawasch Abedjan}
\IEEEauthorblockA{
\textit{TU Berlin \& BIFOLD}\\
Berlin, Germany \\
abedjan@tu-berlin.de}
}

\maketitle
\begin{abstract}
%Data discovery is an iterative process that requires the execution of multiple queries to find the desired tables from large data lakes. 
Most research on data discovery has so far focused on improving individual discovery operators such as join, correlation, or union discovery. However, in practice, a combination of these techniques and their corresponding indexes may be necessary to support arbitrary discovery tasks. 
We propose \system, a comprehensive data discovery system that supports existing operators and enables their flexible pipelining. \system is based on a set of lower-level operators that serve as fundamental building blocks for more complex and sophisticated user tasks.
To reduce the execution runtime of discovery pipelines, we propose a unified index structure and a rule-based optimizer that rewrites  SQL statements into low-level operators when possible. 
We show the superior flexibility and efficiency of our system compared to ad-hoc discovery pipelines and stand-alone solutions.
\end{abstract}

\section{Introduction} \label{sec:introduction}
The potential for large structured data lakes, such as governmental data lakes~\cite{DBLP:conf/aaaiss/DingDGMLMH10, DBLP:journals/corr/abs-2106-09590}, enterprise-level corpora~\cite{DBLP:journals/pvldb/BharadwajGBG21,abs-2402-06282}, and public-access lakes~\cite{Eberius:2015,DBLP:journals/pvldb/CafarellaHWWZ08, hulsebos2021gittables}, has spurred various avenues of research, all aimed at empowering data scientists with the ability to extract further insights for their diverse tasks, such as machine learning (ML)~\cite{DBLP:conf/edbt/EsmailoghliQA21}, exploratory data analysis~\cite{DBLP:conf/sigmod/MiloS20}, and data cleaning~\cite{DBLP:journals/pvldb/MahdaviA20}. 
To support these tasks, various data discovery operatorion have been studied, such as  
\textbf{join discovery}~\cite{DBLP:journals/pvldb/EsmailoghliQA22, DBLP:conf/sigmod/ZhuDNM19, DBLP:journals/tods/XiaoWLYW11, DBLP:conf/icde/XiaoWLS09, DBLP:conf/edbt/VenetisSR12, DBLP:conf/icde/FernandezMNM19, DBLP:journals/pvldb/ZhuNPM16, DBLP:conf/icde/DongT0O21, DBLP:journals/pvldb/CasteloRSBCF21, DBLP:journals/pvldb/SuriIRR21, DBLP:conf/sigmod/Sarawagi04}, %where the goal is to find tables that are joinable with one or multiple columns of a given input table to augment it with new columns; 
\textbf{union search}~\cite{DBLP:journals/pvldb/NargesianZPM18, DBLP:conf/icde/BogatuFP020, DBLP:journals/pvldb/LehmbergB17, DBLP:journals/pvldb/CafarellaHK09, DBLP:journals/pvldb/FanWLZM23, DBLP:journals/pacmmod/KhatiwadaFSCGMR23}, %where the user aims to increase the number of rows in the input table by finding tables that contain complementing information in a similar schema; and
and \textbf{correlation discovery}~\cite{DBLP:conf/edbt/EsmailoghliQA21, DBLP:conf/icde/SantosBMF22, DBLP:conf/sigmod/BecktepeEKA23, DBLP:journals/pvldb/ChepurkoMZFKK20}. 
%, where the objective is to discover joinable tables based on their potential to increase the accuracy of downstream ML models.

Existing research has focused on each operator individually without considering
possible opportunities for combining, pipelining, and optimizing them.
Yet, fulfilling a real-world data discovery task often requires a combined application of these operators.
For instance, in the case of data enrichment for ML, the user needs to use a combination of join, union, and correlation discovery~\cite{DBLP:conf/sigmod/BecktepeEKA23, DBLP:journals/pvldb/NargesianZPM18}. 
To achieve this, we need a unified data discovery system.

\subsection{Requirements for a unified discovery system}
\begin{figure}
    % \vspace{-.15cm}
    \center{\includegraphics[scale=0.20]
          {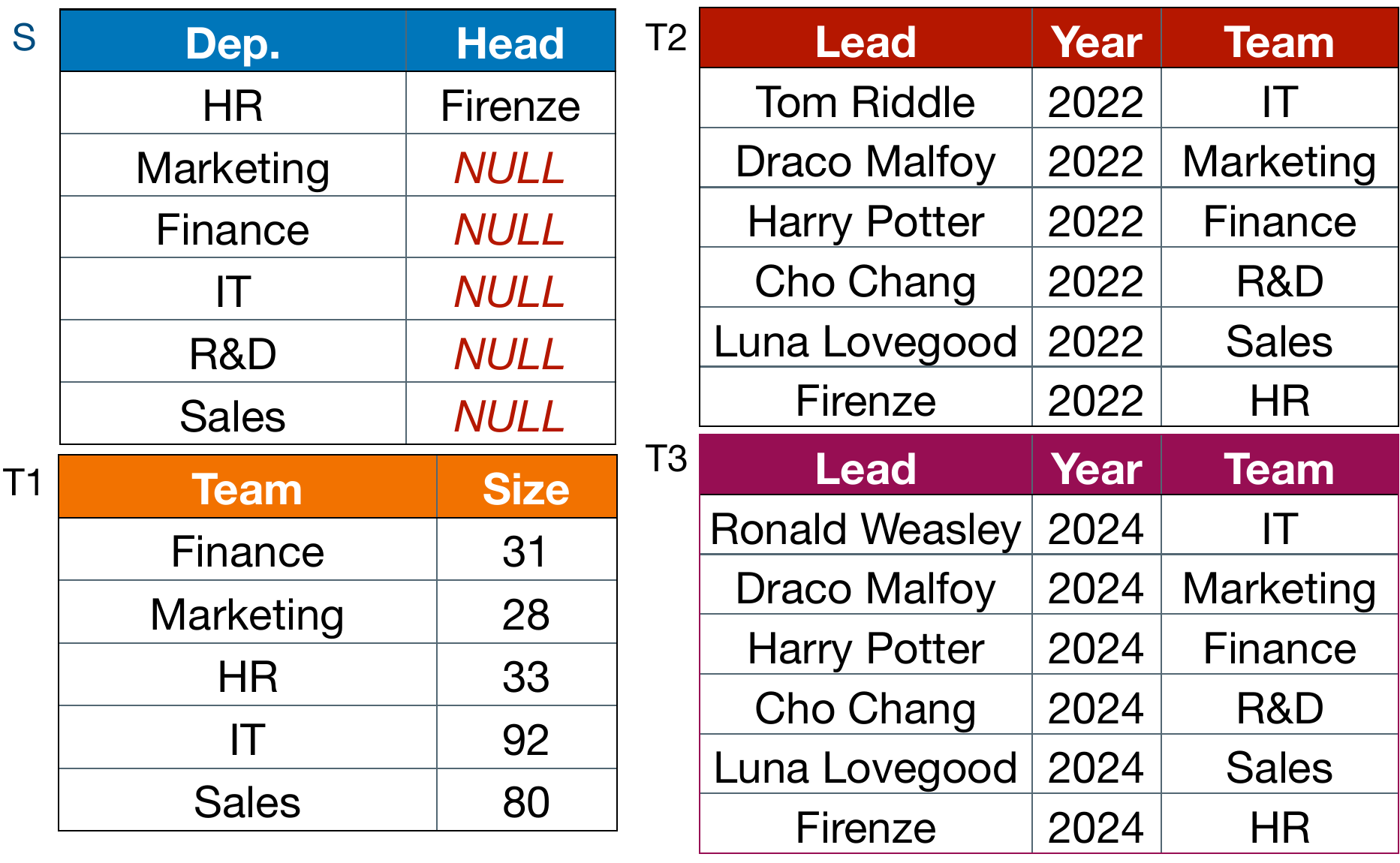}}
    \caption{Discovering up-to-date tables to fill in the NULLs.}
    \label{fig:example}
    \vspace{-.5cm}
\end{figure}

A trivial approach to building such a unified system is to federate all indexes and algorithms via a common interface. Even after overcoming the inherent software and interface heterogeneity among prototypes, the execution of complex data discovery tasks that combines these operators across systems is inefficient and inflexible in terms of both storage and maintenance. This stems from several key challenges:
\newline
\textbf{Lack of Flexibility.} Introducing new criteria or discovery operators is challenging since new indexes and algorithms must be integrated into an already complex system.
\newline
\textbf{Storage and Maintenance Overhead.} Multiple independently developed index structures must be generated and stored. The aggregated system will comprise an incoherent set of potentially redundant structures needing maintenance. %Further, since data lakes can be arbitrarily large, storage costs multiply accordingly.
\newline
\textbf{Lack of Optimization.} The federation of independent operators under the same abstraction does not lend itself to optimization. 
To showcase the operational complexity of a rather simple discovery task, consider following example:

\noindent \textbf{Example 1.}
Fig.~\ref{fig:example} illustrates a user table $S$ along with three tables $T1$, $T2$, and $T3$ from a data lake.
$S$ contains information about departments in a company and the name of their corresponding leaders, i.e.,``Head''.
The user wants to fill in missing values of $S$ but already knows that the head of the IT department, ``Tom Riddle'', has left the company.
Therefore, any table that contains the projection (``IT'',``Tom Riddle'') is outdated. 
Thus, the user defines the data discovery for missing values as follows:
\textit{Find top-$1$ table that contains (\textcolor{teal}{``HR'' and ``Firenze''} in a row and \textcolor{teal}{``HR'', ``Marketing'', ``Finance'', ``IT'', ...} in a column but does not contain rows with \textcolor{red}{``IT'' and ``Tom Riddle''}).}

To solve this task with existing discovery operators, one could use single-column join discovery, such as Josie~\cite{DBLP:conf/sigmod/ZhuDNM19}, to look up overlapping column values, and multi-column join discovery, such as \textsc{Mate}~\cite{DBLP:journals/pvldb/EsmailoghliQA22}, to find overlapping column combination values. 
With \textsc{Mate}, one could find tables containing the projection (``HR'', ``Firenze'') as result set $rs_1$. Also, \textsc{Mate} can be used to identify tables containing the tuple (``IT'', ``Tom Riddle'') as result set $rs_2$. Josie can be employed to find tables that join only on the department values {``HR'', ``Marketing'', ``Finance'', ...} as result set $rs_3$.
Finally, the equation ($(rs_1 \cap rs_3) - rs_2$) is the answer to the user's query.
According to Fig.~\ref{fig:example}, $rs_1 = \{ T2, T3 \}$, $rs_2 = \{ T2 \}$, and $rs_3 = \{ T1, T2, T3 \}$. Thus, the answer to the discovery query is $\{ T3 \}$, which contains amendable values for the ``Head'' of the department column in $S$.\hfill $\bullet$

The example illustrates that running independent discovery redundant and inefficient. 
For each look-up, combinations of the three tables $T1$, $T2$, and $T3$ may be processed.
However, knowing that $T1$ and $T2$ are not desired tables, we could exclude them from follow-up operators: first, by checking all three tables to find those containing (``HR'' and ``Firenze''); then, by focusing only on $T3$ and $T2$ and dropping the outdated $T2$; and finally, by verifying $T3$ for overlapping department names.
Intuitively, we could obtain the same results by resorting to application code. %One would obtain a first set of tables based on the inclusion of {(``HR'', ``Firenze'')} into memory. Then, we would filter tables with outdated entries and search for department-to-head mappings that exist. 
We aim to achieve the same flexibility within a discovery system with a unified language to optimize custom discovery processes. Such abstraction becomes necessary in case of complex queries with a variety of operators and result sets with millions of tables.

\subsection{Contributions}
We introduce \system, a unified system that enables users to create 
% %rjm "arbitrary" is a tall order
complex discovery plans. With no need for compiling, installing, or adapting additional discovery systems, \system streamlines the discovery process.
Given a data discovery plan, a data lake, and a parameter $k$, \system retrieves the top-$k$ tables from the lake that are most relevant to the defined plan. 
Our key contributions are as follows:
\begin{enumerate}[wide, labelwidth=!, labelindent=0pt]
    \item \textbf{Introduction of an abstraction layer:} 
    We design a novel abstraction that unifies the building blocks of different discovery tasks. We propose a set of atomic search operators, which users can declaratively compose to efficiently discover relevant tables. These building-blocks are implemented using primitive database functions, allowing the majority of computations to run within the database.% eliminating the need for users to develop low-level algorithms or indexes.
    \item \textbf{Optimized execution of tasks:}
    We develop a two-phase plan optimizer that combines rule- and cost-based query optimization techniques. It re-orders the execution of operators based on predefined rules, ensuring efficient execution while preserving the intended logic of the discovery task. Furthermore, \system optimizes individual operators by rewriting corresponding queries using intermediate results.
    \item \textbf{Integrated index:} 
    We studied fifty-three index structures from thirty data discovery papers and propose a unified index that efficiently supports the execution of the introduced operators and their compositions. The index is designed as a single database table, enabling the system to additionally benefit from the database-level query optimizations.
    \item \textbf{Simplification of complex tasks using SQL operations:}
    Our proposed index structure allows for the translation of \system's operators into SQL queries. Even complex tasks, such as correlation discovery, are reduced to SQL-based operations, eliminating the need for custom algorithms or external implementations. By integrating multiple operators into declarative pipelines, \system minimizes user effort and simplifies the creation of discovery tasks.
    \item \textbf{User study:} We conducted a user study with data experts to understand their discovery preferences and evaluate the effectiveness of \system in addressing these needs.
\end{enumerate}

\section{Problem Statement}\label{sec:problem_statement}
Given the requirements formulated in the introduction, we can formalize the following problem statement. 
Let $\mathcal{D} = \{T_1, T_2, \dots, T_n\}$ be a data lake, where each \( T_i \) is a table. 
Given a query table $S$ with insufficient data and a discovery plan $P$ by the user, the goal of \system is to enable the formulation of the plan and its optimized compilation for execution. The result of the execution is the subset $ \mathcal{T}' \subseteq \mathcal{D} $ that enriches  $ S $. 
% $ \mathcal{Q}(S) $.
With this functionality in mind, we strive to solve four technical problems (\textbf{Pr.1-4}) to create a unified data discovery system. 

First, the language for defining a plan should support the formulation of the most prominent discovery operations from the literature (\textbf{Pr.1}):

\begin{itemize}[wide, labelwidth=!, labelindent=0pt]
    \item \textbf{Keyword search}: discovered tables must overlap with a given set of keywords.

    \item \textbf{Join discovery}: discovered tables must be joinable with $S$ on one or more columns.

    \item \textbf{Union discovery}: rows in discovered tables must be unionable with rows in $S$.

    \item \textbf{Correlation discovery}: a column in each discovered table must correlate with the user-defined numerical target in $ S $.
\end{itemize}
Second, the system should also support any set-based aggregation of these operation's results (\textbf{Pr.2}).

%To satisfy these two problems, the first goal of this paper is to propose a set of low-level operators, $OP = \{op_1, op_2, ..., op_n\}$, that can compose the aforementioned  and their combinations (\textbf{\textit{G2}}).

% using set combiners, including \textit{intersection}, \textit{union}, \textit{difference}, and \textit{count}, on their result set. 

%Our second goal is to build a unified index to cover all of these operators (\textbf{\textit{G3}}).

Third, we require an index that serves all of the possible discovery operations but is leaner than the ensemble of the required individual indexes.
More formally, assuming that an independent system exists for each of the aforementioned discovery operations $M_{op_i}$, where $1 < i < n$, the storage required for our index to execute any combination of operators in $OP$ must be smaller than
$ \sum_{i=1}^{n} \text{Storage}(M_{op_i})$, where \textit{Storage$(M_{op_i})$} is the storage size required for the state-of-the-art index in $M_{op_i}$ (\textbf{Pr.3}).

Finally, the system should satisfy two runtime constraints. The implementation of each $(M_{op_i})$ in the unified system should yield a comparable runtime while any aggregation and combination of more than two $M_{op_i}$ should yield superior runtime compared to their execution as concatenated black box discovery system. Formally, assuming a discovery plan $P$ is an arbitrary combination of operators in $\mathit{OP}$, i.e, $P \in \mathit{OP}^*$, the execution runtime of $P$ in \system should satisfy
$$\text{Time}(P_{\text{BLEND}}) \leq \sum_{j|op_j \in P} \text{Time}(M_{op_j}).$$

Satisfying the previous constraint should not be a burden on how the user formulates the plan. \system must identify a good (near optimal) ordering of the operations based on the expected runtime and intermediate result size (\textbf{Pr.4})

\section{System Overview} \label{sec:system}
\begin{figure*}
    \center{\includegraphics[scale=0.28]
          {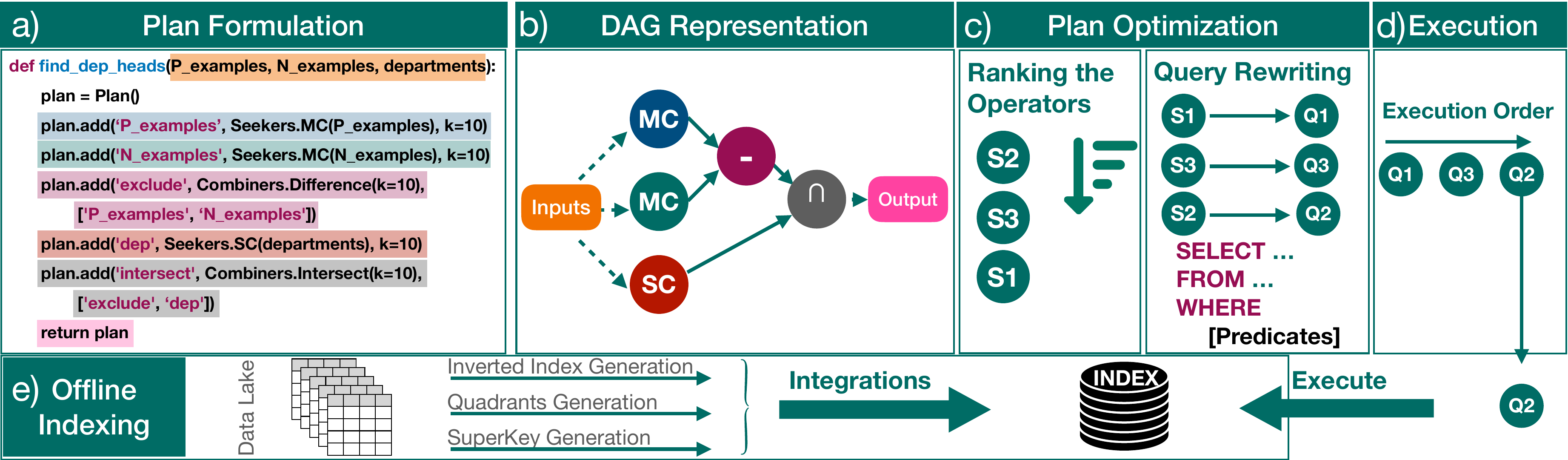}}
          \vspace{-.2cm}
    \caption{\system's architecture.}
    \vspace{-.3cm}
    \label{fig:architecture}
\end{figure*}

Fig.~\ref{fig:architecture} depicts an overview of \system's architecture depicting its online and offline phases. The online phase involves four stages: task definition, plan formulation, plan optimization, and execution. 

First, the user defines their data discovery task using \system's API (Fig.~\ref{fig:architecture}a). 
Here, the plan \textit{find\_dep\_heads} is written for the use case of Fig.~\ref{fig:example}. It finds tables containing a set of positive examples and the given department names but not negative examples.
The main primitives in \system's API are \emph{Plan}, \emph{Seekers}, and \emph{Combiners}.  The latter two,
\emph{Seekers} and \emph{Combiners}, are the lower-level operators in \system, which are implemented in SQL.
%A \texttt{Plan} primitive encapsulates the composition of these operators.
A discovery task is created by instantiating a \emph{Plan} object, to which one can add \emph{Seekers} and \emph{Combiners}. 
%The \texttt{add()} function incorporates a new operator into the plan. 
Seekers are added by specifying an identifier string and parameters, such as input lists and top-$k$. 
%The outputs of these input elements are passed to the new operator for further processing.
%The parameter $k$ specifies the number of tables each operator outputs.
%Fig.~\ref{fig:architecture}a defines the task of data imputation shown in Example $1$, where the 
% input contains a set of positive and negative examples and a set of department names. 
% The 
% The user can specify task-specific parameters, such as the join key for the input table or the target column of a prediction task.
% Initially, the user specifies a data discovery plan using \system's API (Fig.~\ref{fig:architecture}a). 
% The plan includes input data and task-dependent additional parameters, such as the join key for the input table or the target column of a prediction task.
% The plan also consists of a sequence of atomic and simple operators, where the output of one operator serves as the input for the next.
% The user might also pick template plans of common discovery operations, such as join or union discovery. 
\system transforms the plan into a directed acyclic graph of operators, where edges represent the data flow (Fig.~\ref{fig:architecture}b). 
% Fig.~\ref{fig:architecture} depicts the color-coded DAG of the user's plan.
Once the discovery plan is written and the DAG is generated, \system leverages a combination of rule-based and cost-based optimization to re-rank and rewrite operators for faster execution (Fig.~\ref{fig:architecture}c).
The low-level operators either resemble or contain SQL queries on \system's index which is organized as a fact table. 
Based on the cost model, \system rewrites the SQL statements by injecting predicate placeholders enabling the system to take advantage of cardinality estimates of the intermediate results. %The objective of the SQL rewriting is to execute the queries efficiently by increasing the selectivity of the operators.
Once the queries are generated, they are sent to the execution engine to run in the database (Fig.~\ref{fig:architecture}d).
%\system iteratively executes each query and obtains the intermediate results (Fig.~\ref{fig:architecture}d) to update the predicates in subsequent SQL statements to accelerate the plan execution.
In the offline phase, \system builds the necessary index and data structures (Fig.~\ref{fig:architecture}e). The index comprises and unifies three structures: inverted index~\cite{abedjan2015dataxformer}, \textit{super keys}~\cite{DBLP:journals/pvldb/EsmailoghliQA22}, and quadrants~\cite{DBLP:conf/icde/SantosBMF22}. 
%We integrate these structures into a single index, offering a unified interface for operators.

%\answerRthree{In the next sections, we detail the design of individual operators, the construction of the index, the implementation of operators on top of this index, and the policies for execution and optimization of a query plan.}

\section{Operators}\label{sec:operators}
\system's operators are designed to be composable, enabling complex plans, and simple for efficient execution.
\system leverages two sets of operators: seekers and combiners.

\subsection{Seeker operators} \label{sub:seekers}
A seeker receives a set of columns, $Q$, as the input and returns the top-$k$ most relevant tables. 
$k$ is a user-defined parameter. 
We propose four seekers: 
\textit{single-column}, \textit{keyword}, \textit{multi-column}, and \textit{correlation} seekers. 
% These seekers either individually or in arbitrary combinations can cover high-level discovery operators.

% Note that, \system allows the users to define more operators according to their discovery needs.

\subsubsection{Single-Column ($\mathit{SC}$) Seeker} The SC seeker receives an input column $Q$, typically from an input (query) table and returns a list of $k$ tables. These tables contain a column that overlaps the most with $Q$ (sorted by number of overlaps).

% \noindent\textbf{Implementation.} Listing~\ref{sql:sc} shows the implementation of the SC seeker as an SQL query on top of \emph{AllTables}. The query leverages the \texttt{WHERE} clause (Lines~\ref{codeline_sc_select}-\ref{codeline_sc_where}) to find the overlapping columns and returns the top-$k$ tables (Lines~\ref{codeline_sc_group}-\ref{codeline_sc_limit}) sorted based on the column in descending order (Line~\ref{codeline_sc_order}).

% \input{charts_and_tables/code_sc}

\subsubsection{Keyword ($\mathit{KW}$) Seeker} \label{subsubsec:kw_seeker} 
The KW seeker returns the top-$k$ overlapping tables based on a given set of keywords. The main difference between the KW and SC seekers is that the KW seeker measures the overlap throughout entire candidate tables 
%rjm without any concern regarding the value alignments. 
rather than single columns.
% On the other hand, the SC seeker requires the overlap to occur between $Q$ and a single column from the candidate table. 
Returned tables are sorted based on the number of overlaps.
% A standard adaptation of the SC seeker could directly be used for standard keyword search, where the \texttt{ColumnId} can be skipped in the \texttt{GROUP-BY} clause. For brevity, we skip the corresponding SQL code. 

\subsubsection{Multi-Column ($\mathit{MC}$) Seeker} 
The MC seeker receives $Q$ that has multiple columns and discovers the top-$k$ tables in the data lake that the most with the tuples in $Q$. The returned tables are sorted based on the overlap.

\subsubsection{Correlation ($\mathit{C}$) Seeker}
This seeker receives two columns $Q$, $Q=\{Q_j, R\}$, where $Q_j$ is a join key and $R$ is a numerical target column.
% We call the numerical column, the target column.
The correlation seeker returns the top-$k$ tables that join on $Q_j$ and contain a column that correlates the most with $R$, sorted by the correlation coefficient. %The join key defines how the query and candidate tables should be joined to calculate the correlation.

\subsection{Combiner operators} 
Combiners enable the discovery task specific composition of seeker results. Combiners receive a set of table collections, where each collection is the result of a seeker or another combiner. A combiner merges the table collections according to its defined set operation and returns a new set of tables. 
We implemented the most common set operators in \system: \textit{Intersection} ($\cap$), \textit{Union} ($\cup$), \textit{Difference}($\setminus$), and \textit{Counter}. Note that the user can introduce new combiners to the system.
The \textbf{Intersection} and \textbf{Union} combiner receive the output of two or more seekers and returns the intersection/union of the tables, respectively.
The \textbf{Difference} combiner receives only two sets of tables and returns the tables that exist only in the first table set. 
The \textbf{Counter} combiner receives multiple sets of tables, counts the occurrence of the table identifiers, and returns a list of tables sorted in descending order based on their frequencies.

\subsection{Discovery Language Grammar}\label{sec:grammar}
With the given seekers and combiners, 
the grammar of the supported discovery plans in \system can be defined as follows:
\begin{center}
\texttt{expression} ::= \texttt{seeker(Q)} $|$ \texttt{combiner(expression(,expression)+)}
  
\texttt{seeker} ::= \texttt{KW $|$ SC $|$ MC $|$ C}

\texttt{combiner} ::=  $\cap$ $|$ $\cup$ $|$ $\setminus$ $|$ \texttt{Counter}

\texttt{Q} ::= \texttt{keyword} $|$ \texttt{table}
\end{center}
Any \texttt{expression} based on the grammar above can be subject to optimization of \system. Further, it is possible to combine multiple of such expression via native python control flows, each being optimized independently.

%\answer{Before discussing the implementation details of operators introduced in this section, we propose the index structure of our discovery system, \system, in the next section.}
% \subsection{Extensibility Discussion}\label{sub:extend}
% The current implementation of \system is composed of seekers based on exact-matches. However,  \system can be extended with additional operators. To introduce new seekers, the user must create a Python \texttt{class} that inherits from \texttt{Seeker} and implement its interface, which receives a set of columns as input, executes an SQL query, analyzes the results, and outputs a set of table identifiers.
% Similarly, to introduce new combiners, one needs to create a class inheriting from \texttt{Combiner} and implement its interface which receives a collection of sets, each containing a list of table identifiers, and outputs a ranked list of tables.
\section{Index}\label{sec:index}
To efficiently execute the commonly performed discovery tasks, one has to leverage each of the corresponding index structures. 
This approach has several drawbacks with regard to data consistency, application abstraction, and index maintenance. 
To avoid a heterogeneous set of indexes, we make two contributions by first identifying a concise set of necessary indexes and then integrating them within a unified layout: 
\newline
\textbf{I. Identifying relevant indexes.} We studied fifty-three index structures proposed in thirty data discovery papers as illustrated in Table~\ref{tab:paper_to_task_mapping}. Each proposed index addresses a specific data discovery task. Subsequently, we carefully chose three index structures that cover all common discovery tasks and are compatible with each other, then adapted them for \system:
\begin{table}[]
    \scriptsize
    \centering
    \vspace{-.2cm}
    \caption{Mapping between Indexes and data discovery tasks.}
    \label{tab:paper_to_task_mapping}
    
\begin{tabular}{|l||*{5}{c|}} \hline
\makebox[3.2em]{\thead{\textbf{Task} \\ \hline \textbf{Literature} \\ \textbf{References}}}
% \backslashbox[12mm]{\textbf{Index}} {\textbf{Task}}
&\makebox[4.1em]{\thead{Keyword \\ Search \\ \hline \cite{DBLP:conf/sigmod/SarmaFGHLWXY12, DBLP:conf/cikm/ZhangSF21} \\ \cite{DBLP:conf/sigmod/NargesianPZBM20, DBLP:conf/icde/FernandezAKYMS18} \\ \cite{abedjan2015dataxformer, DBLP:journals/pvldb/CafarellaHK09, DBLP:conf/www/BrickleyBN19} }}
&\makebox[8.7em]{\thead{Join Search \\ \hline \cite{abedjan2015dataxformer, DBLP:journals/pvldb/CasteloRSBCF21, DBLP:conf/icde/XiaoWLS09, DBLP:conf/icde/FernandezAKYMS18, DBLP:journals/pvldb/BharadwajGBG21} \\ \cite{DBLP:journals/pvldb/SuriIRR21, DBLP:conf/icde/DongT0O21, DBLP:conf/sigmod/ZhangI20, DBLP:conf/sigmod/Sarawagi04, DBLP:conf/www/BayardoMS07} \\ \cite{DBLP:journals/pvldb/ZhuNPM16, DBLP:journals/tods/XiaoWLYW11, DBLP:journals/pvldb/EsmailoghliQA22, DBLP:conf/sigmod/ZhuDNM19} \\ \cite{DBLP:conf/edbt/VenetisSR12, DBLP:conf/icde/FernandezMNM19, DBLP:conf/sigmod/SarmaFGHLWXY12, DBLP:journals/pvldb/CafarellaHK09}}}
&\makebox[1.5em]{\thead{MC \\ Join \\ Search \\ \hline \cite{DBLP:journals/pvldb/EsmailoghliQA22}}}
&\makebox[4.7em]{\thead{Union Search \\ \hline \cite{DBLP:journals/pvldb/ZhangI19, DBLP:journals/pvldb/NargesianZPM18, DBLP:journals/pvldb/LehmbergB17} \\ \cite{DBLP:journals/pvldb/FanWLZM23, DBLP:conf/sigmod/ZhangI20, DBLP:journals/pvldb/KhatiwadaSGM22} \\ \cite{abedjan2015dataxformer, DBLP:conf/sigmod/SarmaFGHLWXY12, DBLP:conf/cikm/AmsterdamerC21} \\ \cite{DBLP:conf/www/BayardoMS07, DBLP:journals/pvldb/CafarellaHK09, DBLP:conf/icde/BogatuFP020}}}
&\makebox[3em]{\thead{Correlation \\ Search \\ \hline \cite{DBLP:conf/edbt/EsmailoghliQA21, DBLP:conf/icde/SantosBMF22} \\ \cite{DBLP:conf/sigmod/SantosBCMF21}}}\\\hline
\end{tabular}
\vspace{-.4cm}
\end{table}
\begin{enumerate}[wide, labelwidth=!, labelindent=0pt]
    \item To enable content-to-table look-ups, which cover keyword, join, and some union search tasks, we employ ideas from the DataXFormer system~\cite{abedjan2015dataxformer}. It introduced an inverted index, which stores cell values and their corresponding location, i.e., table, column, and row identifiers within a single structure.
    \item \textsc{Mate}~\cite{DBLP:journals/pvldb/EsmailoghliQA22} proposes the XASH index. This hash-based structure %allows for 
    enables fast multi-column join discovery. 
    \item The QCR index~\cite{DBLP:conf/icde/SantosBMF22} calculates the Quadrant Count Ratio statistic (QCR) that approximates linear correlation. Alternatively, one could use \textit{Order Index} proposed in the COCOA paper~\cite{DBLP:conf/edbt/EsmailoghliQA21}, however, this index is tailored for exact and non-linear correlations as opposed to the QCR index, which makes the \textit{Order Index} less efficient. 
    % Unlike the \textit{Order Index}, which requires application code to post-process candidate tables, 
    We modify the QCR index in a way that enables in-DB calculation of the correlation estimation via SQL (see Section~\ref{sub:seekers}).
\end{enumerate}
\textbf{II. Designing the layout.} To provide a unified interface for operators on top of the aforementioned indexes, we propose a simple layout that is easy to maintain, prevents data redundancy, and is storage efficient.
\begin{figure}
    \center{\includegraphics[scale=0.2]
          {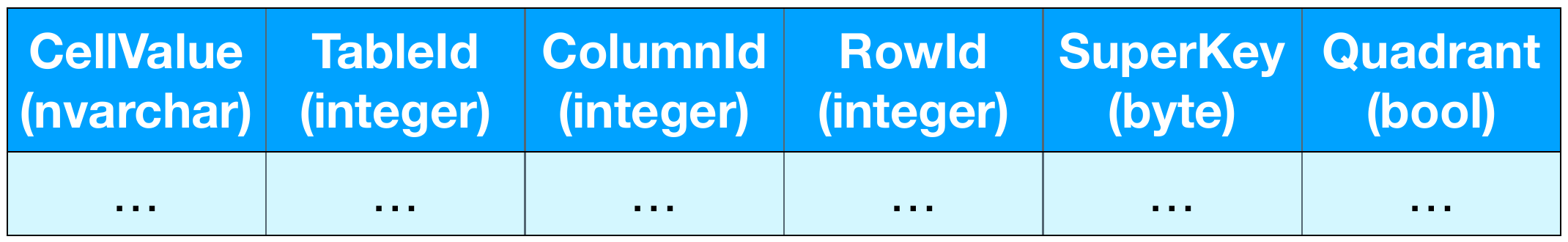}}
    \caption{The lake index is a single relational table: \textit{AllTables}.}
    \vspace{-.5cm}
    \label{fig:index}
\end{figure}
\system serializes these indexes into a database relation called \texttt{AllTables} as shown in Fig.~\ref{fig:index}. Implementing the index as a relation enables \system to benefit from in-DB features, e.g., query optimizations.

\texttt{AllTables} includes four columns from the DataXFormer
index~\cite{abedjan2015dataxformer}: \textit{CellValue}, \textit{TableId}, \textit{ColumnId}, and \textit{RowId}, mapping each value to its locations in the lake. 
To facilitate fast value look-up and table loading, we create two in-database indexes on the \textit{CellValue} and the \textit{TableId} columns, respectively.
An additional column stores the XASH index, i.e., a \textit{super key}, for the given table and row~\cite{DBLP:journals/pvldb/EsmailoghliQA22}. The \textit{super key} aggregates row values into a single hash, enabling efficient discovery of complex multi-column matches.

The original QCR index is not directly compatible with \texttt{AllTables} as it requires the enumeration of column pairs. Santos et al.~\cite{DBLP:conf/icde/SantosBMF22} use the QCR~\cite{holmes2001correlation} statistic to estimate the Pearson correlation between candidate columns and a target. 
To approximate the correlation between two sets of values $X$ and $Y$, QCR divides the observations into four quadrants based on the relative position of the observations compared to their corresponding set averages. 
If a pair of values, $x_i$ and $y_i$, are both greater or both smaller than their corresponding averages, they belong to Quadrant $I$ or $III$, respectively. Otherwise, they are placed in either Quadrant $II$ or $IV$.
The QCR is computed as $QCR = \frac{n_I + n_{III} - n_{II} - n_{IV}}{N}$, where $n_{Quadrant}$ is the number of observations in the quadrant, and $N$ is the number of all observations. 
% The baseline approach embeds QCR information inside the index:
for any two column pairs in a table, where one is categorical and one numerical, the baseline approach hashes each row based on the categorical value and the quadrant entry of the numerical value. The smallest $h$ hashes are stored inside the index. During retrieval, the same hash is generated for the join and the target columns in $Q$ and the minimal values are matched against the entries in the index.

Using our index structure, we can calculate the correlation inside the database avoiding the need for further data loading and serialization to the application level.
To align the original index with our design, we propose a novel representation that only requires one additional \texttt{Quadrant} column that stores a Boolean per cell. 
\textit{Quadrant} is set to \texttt{1} if the cell value is larger than or equal to the column average, and \texttt{0} otherwise. For Non-numerical data points \textit{Quadrant} is set to \textit{Null}. 
In contrast to the baseline, we invoke the sample size $h$ while querying. With this approach, we do not constraint the comparison to $h$ hash values that are smallest but simply sample $h$ hash values at random. 
This has the benefit that the hash size can be dynamically chosen, which in the original approach requires re-indexing the lake. 
% Thus instead of focusing of comparing hash values that are smallest, we assume a random set of hash values. 

\section{Seeker Implementations}\label{section:seeker_implementation}
We now discuss operator implementations by detailing how operators use the index to efficiently return the top-$k$ tables.

\noindent\textbf{SC Seeker.}
Listing~\ref{sql:sc} shows the implementation of the SC seeker as an SQL query on top of \emph{AllTables}. The query leverages the \texttt{WHERE} clause (Lines~\ref{codeline_sc_select}-\ref{codeline_sc_where}) to find the overlapping columns and returns the top-$k$ tables (Lines~\ref{codeline_sc_group}-\ref{codeline_sc_limit}) sorted based on the column in descending order (Line~\ref{codeline_sc_order}).

\begin{lstlisting}[language=SQL, caption={SQL implementation of the SC seeker.}, label={sql:sc}, escapechar=|]
SELECT TableId FROM AllTables |\label{codeline_sc_select}|
WHERE CellValue IN (Q) |\label{codeline_sc_where}|
GROUP BY TableId, ColumnId |\label{codeline_sc_group}|
ORDER BY COUNT(DISTINCT CellValue) DESC |\label{codeline_sc_order}|
LIMIT K;    |\label{codeline_sc_limit}|
\end{lstlisting} 

\noindent\textbf{KW Seeker.}
A standard adaptation of the SC seeker could directly be used for standard keyword search, where the \texttt{ColumnId} can be skipped in the \texttt{GROUP-BY} clause. For brevity, we skip the corresponding SQL code. 

\noindent\textbf{MC Seeker.}
The MC seeker requires not only identifying tables containing values from all input columns but also ensuring consistent value alignment.
This means that the values from the same row in $Q$ must occur in the same row in the candidate tables. 
MATE~\cite{DBLP:journals/pvldb/EsmailoghliQA22} ensures this value alignment by representing table rows via an aggregated hash, i.e., \textit{super key}, which serves as a bloom filter to prune non-joinable rows without checking the individual values and their alignment. 

First, \textsc{Mate} uses an SQL query to only detect candidate table rows with any arbitrary combination of values from $Q$. 
The SQL query is shown in Listing~\ref{sql:mc_overlap}. Note that the listing only shows the generated query for two columns but \system can generate the query for any number of columns.

% \begin{lstlisting}[language=SQL, caption={SQL implementation of the MC overlap seeker.}, label={sql:mc_overlap}, escapechar=|]
% SELECT output1.TableId FROM    |\label{codeline_mc_select}|
%     SELECT TableId, RowId FROM AllTables     |\label{codeline_mc_innerselect1}|
%     WHERE CellValue IN (Q1)) AS output1     |\label{codeline_mc_where1}|
%     INNER JOIN     |\label{codeline_mc_join}|
%     (SELECT TableId, RowId FROM AllTables      |\label{codeline_mc_innerselect2}|
%     WHERE CellValue IN (Q2)) AS output2     |\label{codeline_mc_where1}|
%     ON output1.TableId = output2.TableId      |\label{codeline_mc_on}|
%     AND output1.RowId = output2.RowId      |\label{codeline_mc_and}|
% GROUP BY output1.TableId      |\label{codeline_mc_group}|
% ORDER BY COUNT(*) DESC      |\label{codeline_mc_order}|
% LIMIT k;      |\label{codeline_mc_limit}|
% \end{lstlisting} 

\begin{lstlisting}[language=SQL, caption={The first phase of the MC seeker.}, label={sql:mc_overlap}, escapechar=|]
SELECT * FROM    |\label{codeline_mc_select}|
    (SELECT * FROM AllTables     |\label{codeline_mc_innerselect1}|
    WHERE CellValue IN (Q1)) AS Q1_index_hits     |\label{codeline_mc_where1}|
    INNER JOIN     |\label{codeline_mc_join}|
    (SELECT * FROM AllTables  |\label{codeline_mc_innerselect2}|
    WHERE CellValue IN (Q2)) AS Q2_index_hits |\label{codeline_mc_where1}|
    ON Q1_index_hits.TableId = Q2_index_hits.TableId|\label{codeline_mc_on}|
    AND Q1_index_hits.RowId = Q2_index_hits.RowId |\label{codeline_mc_and}|
\end{lstlisting} 

The query fetches index rows, including hash values, where the values from mutually exclusive columns of the same table appear in the same set of candidate table rows.
The exact alignment of actual tuples is validated on the application level.
% and phase 2 as shown in Fig.~\ref{fig:mc_architecture}. 
Then, it uses the \textit{super key} of each row to filter rows with wrong alignment of value combinations from $Q$ and validates the remaining rows through exact match validation.

\begin{lstlisting}[language=SQL, caption={SQL implementation of the correlation seeker.}, label={sql:dependency}, escapechar=|]
SELECT Keys.TableId FROM |\label{codeline_correlation_select}|
(SELECT * FROM AllTables WHERE rowid < h AND C_Values IN ($Q_j$)) keys |\label{codeline_sub_query_keys}|
INNER JOIN  |\label{codeline_innerjoin}|
(SELECT * FROM AllTables WHERE rowid < h AND Quadrant is not NULL) nums |\label{codeline_sub_query_nums}|
ON keys.TableId = nums.TableId AND keys.RowId = nums.RowId |\label{codeline_correlation_on}|
GROUP BY keys.TableId, nums.ColumnId, keys.ColumnId |\label{codeline_correlation_group}|
ORDER BY ABS($SCORE$) DESC     |\label{codeline_correlation_order}|
LIMIT k;     |\label{codeline_correlation_limit}|
\end{lstlisting}

\noindent\textbf{Correlation Seeker.}
Listing~\ref{sql:dependency} shows the implementation of the correlation seeker.
First, the query fetches value pairs, where the first value is from $Q_j$ and the second one is a value from a numerical column (Lines~\ref{codeline_correlation_select}-\ref{codeline_correlation_on}). This is done through an \texttt{INNER JOIN}.
To ensure that the second value is numerical, we checks if \textit{Quadrant} is non-NULL.
Both \texttt{WHERE} clauses in the sub-queries (Line~\ref{codeline_sub_query_nums}) sample $h$ rows for correlation calculation as described in Section~\ref{sec:index}. 
The result of the join is grouped based on the combination of the candidate table, its join key, and numerical column (Line~\ref{codeline_correlation_group}).
Then, we sort each triplet by their QCR score (Line~\ref{codeline_correlation_order}) and return the top-$k$ tables (Line~\ref{codeline_correlation_limit}).
% The QCR score is calculated as $\frac{n_I + n_{III} - n_{II} - n_{IV}}{N}$.  
As $n_{II} + n_{IV} = N - (n_{I} + n_{III})$, we calculate QCR  as $\frac{2 \times (n_I + n_{III}) - N}{N}$, which in SQL is implemented as:
\begin{lstlisting}[language=SQL, escapechar=|]
(2|$\times$| SUM(((keys.C_Values IN ($k_0$) AND nums.Quadrant = 0) OR (keys.tokenized IN ($k_1$) AND nums.Quadrant = 1))::int)- COUNT(*)) / Count(*)
\end{lstlisting} 
\textit{\$k\_0\$} are the input join keys whose corresponding target value is below the average and \textit{\$k\_1\$} represent join keys whose corresponding target values are greater or equal to the average of the target column. 
The splitting happens before invoking the query while parsing the input table.
% A value pair from the input target and a candidate numerical column belongs to the first or third quadrants if both values are either below or above their corresponding averages.

\begin{sloppypar}
Our implementation has three advantages over the original implementation~\cite{DBLP:conf/icde/SantosBMF22}:
\textbf{i)} We calculate the absolute value for QCR in one run, without the need to calculate positive and negative correlations twice.
\textbf{ii)} Our implementation supports numerical join keys as opposed to the baseline.
\textbf{iii)} Our index does not require quadratic complexity to store the QCR index.
\end{sloppypar}
\section{Defining and Optimizing Discovery Plans}\label{sec:pipeline}
%In the previous section, we introduced the seekers and combiners
% , along with their implementation 
%on top of our proposed index.
% and database schema
By defining seekers and connecting them with combiners, users can declaratively define novel discovery plans. 
%a declarative API that enables users to generate their own tailored data discovery plans.
%As these user-defined discovery plans can contain any combinations of seekers and combiners
Hence, there will be potentials for optimization through reordering of the operators. 
To this end, we propose an optimization approach to execute each plan efficiently.

\subsection{Composing discovery tasks with \system}
\label{sub:composing_discovery_task}
We now explain how to assemble a task using \system's API.
% by example tasks from literature and one complex task. 
%rjm I suggest de-emphasizing the language
%The query language in 
\system covers a wide range of discovery tasks:

\noindent\textbf{Simple tasks} can contain a single seeker.
A user can perform single-column join~\cite{DBLP:conf/sigmod/ZhuDNM19}, Multi-column join~\cite{DBLP:journals/pvldb/EsmailoghliQA22}, and Correlation search~\cite{DBLP:conf/icde/SantosBMF22, DBLP:conf/sigmod/SantosBCMF21, DBLP:conf/edbt/EsmailoghliQA21} tasks, simply by using individual seekers dedicated to each task.
%rjm —SC, MC, and correlation, respectively.}

\noindent\textbf{Union search} aims at discovering tables that can be unioned with the query table, requires a combination of seekers and combiners.
The Union discovery plan in \system comprises multiple SC seekers, one per input table column, and one \textit{Counter} combiner that aggregates the results of these seekers.
We choose a higher top-$k$ limit for individual seekers than the combiner to include tables become relevant when multiple columns are considered in combination~\cite{DBLP:journals/pvldb/NargesianZPM18}. 
% In the end, the combiner only returns the $k$ most frequent tables.
% In our experiments, we use $k' = 100$.

With the 
%rhn so far mentioned 
operators described above, \system is already more expressive than any state-of-the-art system. Furthermore, it allows for any combination of these operators with combiners. 
% Using \system, the user can form any query requires exact overlap similarity or correlation constraint. In particular, \textit{SC}, \textit{MC}, and \textit{KW} seekers enable vertical, horizontal, and arbitrarily-aligned overlaps and the \textit{C} operator incorporates correlations.

\noindent\textbf{Multi-objective data discovery} is the combined task of finding tables from the lake that can enrich the query table with additional rows and columns as well as filling the empty cells.
Listing~\ref{code:augmentation_by_example} shows the plan definition for this task using \system's API. The task is comprised of keyword search, union search, data imputation, and correlation discovery.

We initialize the plan object in Line~\ref{codeline_augmentation_planclass}.
In Line~\ref{codeline_kw_addkw} we create the keyword seeker and add it to the plan.
Next, we add the union search sub-plan. We add one SC seeker per input column (Lines~\ref{codeline_union_for} - \ref{codeline_union_addoverlap}), followed by the \textit{Counter} combiner (Line~\ref{codeline_union_add_combiner}).

% To take advantage of the \textit{Counter} combiner, we use a higher $k$ limit for individual SC seekers ($k'$). This helps discover tables that may not always appear among the top-$k$ results for individual columns, yet they are relevant to more columns in the query table. In the end, the combiner only returns the $k$ most frequent tables.
% In our experiments, we use $k' = 100$.

\begin{figure}
    \center{\includegraphics[scale=0.4]
          {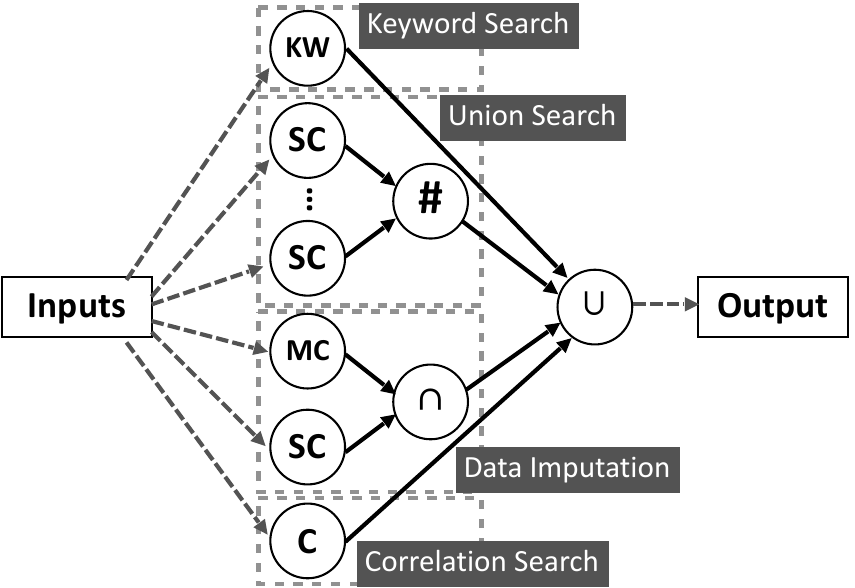}}
    \caption{Multi-objective discovery plan.}
    \label{fig:pipeline_example}
    \vspace{-.4cm}
\end{figure}

Next, we add a data imputation sub-plan. To find tables with complete rows, we use an MC seeker (Line~\ref{codeline_augmentation_addmc}) and for rows where only values of a single column exist, we use an SC seeker (Line~\ref{codeline_augmentation_addoverlap}).
We combine the results of both seeker types in  Line~\ref{codeline_augmentation_combiner} with an intersection combiner. 
Similarly, we create the correlation seeker (Line~\ref{codeline_corr_add}). 
Finally, we aggregate the results of each sub-plans via a \textit{Union} combiner (Line~\ref{codeline_union_combiner}).
% , representing the final state of the plan. 
Fig.~\ref{fig:pipeline_example} illustrates the DAG representation of this plan. We introduce more complex tasks in Section~\ref{sec:experiments}.

\begin{lstlisting}[language=Python, caption={Building search plan in \system.}, label={code:augmentation_by_example}, escapechar=|]
def build_search_plan(keywords, examples, queries):
  plan = Plan()|\label{codeline_augmentation_planclass}|
  # Keyword Search
  plan.add('kw', Seekers.KW(keywords, k=10))|\label{codeline_kw_addkw}|
  # Union Search
  for clm in examples.columns:|\label{codeline_union_for}|
    plan.add(clm, Seekers.SC(examples[clm], k=100))|\label{codeline_union_addoverlap}|
  plan.add('counter', Combiners.Counter(k=10), examples.columns)|\label{codeline_union_add_combiner}|
  # Data Imputation
  plan.add('examples', Seekers.MC(examples, k=10))|\label{codeline_augmentation_addmc}|
  plan.add('query', Seekers.SC(queries, k=10))|\label{codeline_augmentation_addoverlap}|
  plan.add('intersection', Combiners.Intersect(k=10), ['examples', 'query'])|\label{codeline_augmentation_combiner}|
  # Correlation Search
  plan.add('correlation', Seekers.Correlation(examples.joinkey, examples.target, k=10))|\label{codeline_corr_add}|
  # Results Aggregation
  plan.add('union', Combiners.Union(k=40), ['kw', 'counter', 'intersection', 'correleation'])|\label{codeline_union_combiner}|
  return plan
\end{lstlisting}

\subsection{Plan Optimization and Execution}\label{sub:optimization}
We describe how \system optimizes and executes a user-defined discovery plan to solve \textbf{Pr4} as discussed in Section~\ref{sec:problem_statement}.
%\subsubsection{Optimization Objective}
%Given a discovery plan, the goal of the optimizer is to execute the plan as efficiently as possible. 
%The optimizer must maintain the logical sequence of the operators, i.e., the structure of the plan, avoiding any logical errors or inconsistencies that might arise from reordering or accelerating seekers. 
%In particular, the optimizer should account for dependencies between operators, rewriting opportunities, and operator push-downs to enhance performance. 

\subsubsection{Naive Optimization}
To run a plan, the system needs to execute a sequence of SQL queries generated for each seeker and use combiners to aggregate their results.
There are two naive solutions to streamline the queries: 
\begin{itemize}[leftmargin=*]
    \item Executing concurrent seekers independently and aggregating the results from each SQL statement or
    \item Pushing down the combiners to the database.
    % \item Combining SQL queries using database operators, e.g., \texttt{UNION, INTERSECT, EXCEPT}, into one statement that is to be optimized by the underlying database.
\end{itemize}

The first approach does not use inter-query dependencies to speed up execution.
As depicted for the data imputation subplan in Fig.~\ref{fig:pipeline_example}, the results of seekers could be intersected, which provides the option of reusing the results of one seeker to limit the search space of a concurrent seeker.
The second approach also cannot benefit from such optimization potentials, as SQL \emph{set} operators inside the database are oblivious to the size of subqueries that are its operands. 
They optimize each subquery in isolation.
Moreover, some seekers, such as the MC seekers, can include additional application-level code.
To address these problems, \system employs query rewriting.

\subsubsection{The Optimizer in \system}
% \textbf{The Optimizer in \system.}
% The optimizer in \system receives a discovery plan as input and returns a sequence of SQL statements to be executed in the database.
The optimizer in \system receives a discovery plan as input and returns a high-level execution plan as output.
This execution plan contains instructions regarding the ranked sequence of SQL statements, how intermediate results can expedite the next query for faster in-DB execution, and how to apply query re-writings to assist the lower-level DBMS optimizer.

% We use a hybrid approach to optimize discovery plans. 
The heuristic of the optimizer is that executing the faster seeker first and using the obtained intermediate results to limit the search space for the next seeker increases the efficiency. 
According to our experiments, this heuristic is effective in $96\%$ of the $4000$ randomly sampled cases.
The optimizer undergoes four steps: Execution Group (EG) identification, EG ordering, operator ranking, and query rewriting.
% operator ranking, and query rewriting. 

% \subsubsection{EG Identification}
\noindent \textbf{EG identification.} 
To ensure that the optimizer preserves the semantic of the original plan, we discover isomorphic plans that deliver the same output.
To this end, the optimizer identifies seekers that can be re-ordered without changing the plan's output. 
Note that although seekers that are connected to the same combiner can be reordered, we only change the order of seekers connected to the \textit{Intersection} combiner. The remaining combiners are either non-commutative, i.e., \textit{Difference}, or they cannot benefit from the operator re-ordering, i.e., \textit{Union} and \textit{Count}.
% In fact, seekers that are connected to the same combiner can be reordered (Except the \textit{Difference} combiner). \answerRthree{Note that union, intersection, and counter are commutative and allow a change of order without changing the results.}
The optimizer traverses the DAG of the discovery plan and identifies all of such so-called execution groups (EG) and builds a hyper-DAG on top of the original DAG. 

% \subsubsection{EG Selection}
\noindent \textbf{EG ordering.} 
Once the EGs are discovered, the optimizer follows a topological order based on the dependency of the EGs and processes one EG after another. An EG is selected only after all its dependencies are executed.
Once the next EG is selected for execution, \system optimizes the sub-plan of the EG before executing its seekers in the database.

\noindent \textbf{Operator ranking.}
Following the heuristic of executing faster seekers first, \system ranks the seekers based on their expected runtime, which depends on three crucial parameters:
% EG optimization comprises the reordering of operators within an EG and the rewriting of the corresponding SQL queries.
% As the data imputation EG in Fig.~\ref{fig:pipeline_example} depicts, one can use the output result of a seeker to limit the search space of a concurrent seeker. 
% Our experiments show that running the faster seeker first, results in a more efficient plan in $96\%$ of the cases.
% The runtime of seekers depends on three crucial parameters:
\begin{enumerate}[leftmargin=*]
\item the \textbf{seeker type}, as each type has a different runtime complexity which can be determined through apriori analysis,
    \item the \textbf{number of rows projected by $Q$}, and
    \item the \textbf{frequency of values from $Q$ in the database}.
\end{enumerate}
Using these parameters, we build a two-step re-ordering approach: rule-based ranking and learning-based cost estimation.
Similar to logical optimizers in databases, we first reorder seekers based on types. 
According to the implementation details of the operators in Section~\ref{section:seeker_implementation}, the $\mathit{KW}$ operator has the lowest complexity among operators. 
It requires a single index scan, during which \textit{Cell Value}s are compared to $Q$. If there are $n$ elements in the index and $|Q|$ is the number of keywords $Q_{\text{KW}}$, $\mathit{KW}$ complexity is $O(n \cdot |Q|)$.
Although, $\mathit{SC}$ yields the same complexity of $O(n \cdot |Q|)$, the input size $|Q_{\text{SC}}|$ for $\mathit{SC}$ is a column compared to a few keywords in $\mathit{KW}$, making $|Q_{\text{SC}}|\geq |Q_{\text{KW}}|$.
The correlation operator ($\mathit{C}$) requires two index scans, one to find overlapping columns with join key leading to the same complexity as $\mathit{SC}$ ($O(n \cdot |Q|)$), and one scan to find candidate numerical columns $O(n)$. The operator joins the results of these scans, which is linear for hashjoins, resulting in overall complexity of $O(3 \cdot n \cdot |Q|)$. 
% It requires a single index scan and if there are $n$ elements in the index, KW complexity is $O(n+g_t)$, where $g_t$ is the number of \texttt{COUNT()} statements obtained by grouping the tuples by the number of tables. The complexity of SC is similar to that of KW except the required number of \texttt{COUNT()} operators, which is $g_c$. We know that the number of all columns is higher or equal to the number of tables ($g_c \geq g_t$), making SC slower than KW. The correlation operator (C) requires two index scans ($2 \cdot n$) and a join operation ($2 \cdot n$), leading to a higher complexity of $O(4 \cdot n)$. Note that $n >> g_c$.
Finally, the $\mathit{MC}$ operator requires $x$ index scans, one for each input column, and $x-1$ hashjoins to combine the results, therefore, it has the highest complexity of $O(x \cdot n \cdot |Q| + (x - 1) \cdot 2 n)$, where $x$ is the number columns in the input composite key and $x \geq 2$. Furthermore, $\mathit{MC}$ has additional application-level code, adding additional overhead.
Hence, we can summarize the execution order with the following rules:
\begin{itemize}[leftmargin=*]
    \item \textbf{Rule 1.} The keyword operator always executes first.
    \item \textbf{Rule 2.} The MC seeker always executes last.
    \item \textbf{Rule 3.} SC is prioritized over C.
\end{itemize}
%The rule-based re-ordering is fast and accurate as different seeker types have individual runtime bounds. 
We implement the rules as \texttt{if-else} clauses in the optimizer.

% In \system, the rule-based optimizer always prioritizes SC seeker over C seeker and C seeker over MC.

\noindent \textbf{Learning-based cost estimation.}
For seekers of the same type, the runtime depends on $Q$ and the frequency of the input values in the database. 
%It is important to mention that $Q$'s size also includes the number of columns in $Q$ for the MC seeker.
We train a regression model per seeker type to estimate the relative runtime. 
%As the regressor-based optimizer ranks the seekers of the same type, the runtime only depends on the input, $Q$.
We encode the input using three features: \textit{cardinality of $Q$}, \textit{the number of columns involved in $Q$}, and \textit{the average frequency of values from $Q$ in the database}. 
For MC, the average frequency feature is calculated by multiplying the average frequencies of columns in $Q$.
This is because the SQL query for MC performs a join between the obtained index elements for each column. 
%The target value to learn is the runtime.

For training each seeker type regression model, we randomly sample $1000$ input $Q$s from the Gittables data lake (See Table~\ref{tab:datalakes}).
Then, we execute the seekers independently and measure the execution runtime to collect the ground truth.
%The seeker execution includes SQL query generation based on the input, SQL query execution in the database, and application code execution.
% , and split the obtained results into train and test sets with $80/20$ split ratio. 
Training occurs offline during deployment, but predicting and ordering seekers is part of the online optimization process. 
The training time on the largest lake was $60$ seconds. Generally, one could assume that the training is transferable but it is advisable to run the training module once upon lake installation.
% In Section~\ref{sub:optimizer_evaluation}, we show that our re-ordering technique correctly selects the faster seeker in $84\%$ of cases.
%To efficiently obtain the frequency information for values of an input table, we leverage an in-memory dictionary, which maps cell values in the data lake to their corresponding frequencies.

\noindent \textbf{Query rewriting.} 
Once the seekers are ranked, %they can be executed inside the database in that order. 
\system starts with the fastest seeker and uses the intermediate results of each executed seeker to speed up the next one. 
\system rewrites the SQL queries by adding combiner-dependent predicates that check  the intermediate results of the previous step:
% for reducing the result set.
% The rewriting depends on the combiner
% for all operators except the first one. It adds predicates that use the intermediate results of previous queries for reducing the result set.
%The rewriting depends on the combiner:
\begin{itemize}[leftmargin=*]
    \item \textit{Intersection}: \texttt{WHERE TableId IN (...)}
    \item \textit{Difference}: \texttt{WHERE TableId NOT IN (...)}
    \item \textit{Count}: \texttt{GROUP BY TableId ORDER BY COUNT(*)}
    \item \textit{Union}: No rewriting
\end{itemize}
%For \textit{Intersection} and \textit{Difference}, \system leverages \texttt{IN} and \texttt{NOT IN} clauses respectively in the \texttt{WHERE} statement. 
For the \textit{Count} combiner a \texttt{GROUP BY} is needed to count the occurrence of tables and descendingly order the results by their frequencies. For \textit{Union} combiners a rewriting is not needed.
This re-writings are implemented via replacing pre-defined placeholders in each seekers' default SQL statement.

% \subsubsection{Query Rewriting}

% \subsubsection{Example}
\noindent \textbf{Example 2.}
Consider the data imputation sub-plan in Fig.~\ref{fig:pipeline_example}.
First, the optimizer identifies independent seekers. Both the \textit{MC} and the \textit{SC} seekers are connected to the same \textit{Intersection} combiner, thus, they form an EG.
% because they are connected to the same \textit{Intersection} combiner. 
As \textit{SC} is faster (\textbf{Rule 2}), \system prioritizes its execution. The \textit{MC} seeker is rewritten using the table identifiers obtained by the \textit{SC} seeker as shown in Listing~\ref{sql:mc_overlap} via a limiting predicate:
\begin{lstlisting}[language=SQL, escapechar=|]
WHERE Q1_index_hits.TableId IN (|$IR_{SC}$|).
\end{lstlisting} 
% shown in Listing~\ref{sql:rewriting}. 
% \input{charts_and_tables/code_rewriting}
\texttt{$IR_{SC}$} is the list of table identifiers discovered by the \textit{SC} seeker.
After query rewriting, the operator is executed. \hfill $\bullet$
% For each SQL query, the predicate placeholder is replaced by intermediate results derived from the preceding query, followed by the execution of the current SQL query. \hfill $\bullet$

\noindent \textbf{In-DB Optimization} \system’s optimizer acts before the in-DB optimizers, it complements in-DB optimizers by revising discovery plans and rewriting queries before execution.
%Thus, our approach complements in-DB optimizers.% by enhancing efficiency through semantic knowledge on the discovery plan. 
% The in-DB optimizer then optimizes the generated query using its cost model.

\begin{theorem}\label{theorem_optimizer}
Given the proposed combiners and seekers, \system's optimizer does not alter the output of the query.
\end{theorem}
\begin{proof}
The query is subject to operator ranking and rewriting. Operator ranking only applies to seekers connected by an \textit{Intersection} combiner, which is commutative, so the order of execution is irrelevant.
The rewriting rule for \textit{Intersection} ensures that only results from intersected seekers are included, preserving correctness. For \textit{Difference}, the re-writing rule filters out the previously discovered tables, aligning with the difference operation. The rule used for the \textit{Count} combiner only counts the frequency of table identifiers and does not alter the underlying data. \textit{Union} requires no rewriting.
Thus, the optimizer preserves the correctness of queries.
\end{proof}

\section{Experiments}\label{sec:experiments}
We carried out a series of experiments to compare \system's efficiency and effectiveness for executing complex as well as simple discovery tasks with individual stand-alone techniques and their ad-hoc compositions. We also conducted a user study on the needs and preferences of data practitioners.
%We answer these questions:
%{\em To what extent can \system reduce the deployment costs of ad-hoc discovery tasks?}
%{\em What are the trade-offs between a generalizable solution and complex stand-alone baselines and how do individual operators perform?}
%{\em What is the influence of the index on the efficiency of the seekers?}
%{\em What are the discovery preferences of data practitioners, and to which extent does \system meet those?}

\subsection{Experimental Setup}
We utilized ten data lakes to compare each higher-level operator to its baseline on the benchmarks from corresponding original papers.
Table~\ref{tab:datalakes} provides details about each of these lakes. 
% The DWTC~\cite{dwtc} is the largest data lake with over $145M$ tables.
% \answer{We also use labeled web table lake from the Lakebench paper~\cite{DBLP:journals/pvldb/DengCCYCYSWLCJZJZWYWT24}.}
% We crawled the German open data lake~\cite{germanGovData} with over $17k$ tables. Gittables~\cite{hulsebos2021gittables} contains over $1.5M$ tables crawled from GitHub repositories.
WDC~\cite{wdcLake} and Canada-US-UK Open Data~\cite{DBLP:conf/sigmod/ZhuDNM19} are used to benchmark Josie and only contain column identifiers, therefore, the number of tables is not known.
% TUS and SANTOS are data lakes for union search benchmarks~\cite{DBLP:journals/pvldb/NargesianZPM18,DBLP:journals/pacmmod/KhatiwadaFSCGMR23}.
% Finally, the New York City open data~\cite{nycLake} has been used in the correlation search baseline paper~\cite{DBLP:conf/icde/SantosBMF22}.

\begin{table}[]
    \footnotesize
    \centering
    \vspace{-.2cm}
    \caption{Data lakes used in the experiments.}
    \label{tab:datalakes}
\begin{tabular}{l|l|l|l}
\toprule
\textit{\textbf{Data lake}} & \textit{\textbf{Tables}} & \textit{\textbf{Columns}} & \textit{\textbf{Rows}} \\ \toprule
DWTC~\cite{dwtc}           & 145M      &  760M  & 1.5B \\
Lakebench Webtable Large~\cite{DBLP:journals/pvldb/DengCCYCYSWLCJZJZWYWT24} & 2.8M    &  14.8M  & 63.7M \\
Gittables~\cite{hulsebos2021gittables}             & 1.5M                      &  16.8M  & 345M \\
German Open Data~\cite{germanGovData}             & 17,144                      & 440K   & 62M       \\
WDC~\cite{wdcLake}          & -                      & 163M   &    1.6B     \\
Canada, US, and UK Open Data~\cite{DBLP:conf/sigmod/ZhuDNM19} & -                      &  745K  & 1.1B \\
TUS~\cite{DBLP:journals/pvldb/NargesianZPM18}          & 1,530                   & 14.8K & 6.8M \\
TUS Large~\cite{DBLP:journals/pvldb/NargesianZPM18} & 5,043 & 55K & 9.6M \\
SANTOS~\cite{DBLP:journals/pacmmod/KhatiwadaFSCGMR23}        & 550                           & 6,322  & 3.8M \\
SANTOS Large~\cite{DBLP:journals/pacmmod/KhatiwadaFSCGMR23}            & 11,090                      & 121K  & 85M \\
NYC open data~\cite{nycLake}          & 1,063                      & 16K   & 290M 
% Synthetic & 500K& 1M&5B
\end{tabular}
\vspace{-.4cm}
\end{table}
 
\subsubsection{\system Implementation and Deployment}
We evaluate \system on top of two different database engines, PostgreSQL (row store) and a commercial column store. If not explicitly mentioned, both \system and the baselines run on the commercial column store. Also, the default value for $k$ is $10$.
We run all experiments on a machine with $64$ processing cores, $512$GB of main memory, and $10$TB SSD storage. The code for \system is available on our GitHub repository\footnote{\url{https://github.com/LUH-DBS/Blend}}.

\subsubsection{Operators And Baselines}
Since we propose a holistic system, we also focus on holistic evaluation. This indicates that we cannot apply the same level of ablation as standalone algorithm papers.  Rather,
we evaluate \system on well-known tasks: Single- and multi-column join search, union search, correlation search, and a set of more complex tasks.
As there is not a single discovery system to cover all of the discovery operators, we compare to individual state-of-the-art baselines per task.

\noindent \textbf{Single-column join search.} 
We use Josie~\cite{DBLP:conf/sigmod/ZhuDNM19} as the state-of-the-art single-column join discovery baseline.
Note that we do not evaluate keyword search because it is an attribute-agnostic version of the join search.
\textbf{Multi-column join search.}
We use \textsc{Mate}~\cite{DBLP:journals/pvldb/EsmailoghliQA22}, the only approach to discover multi-column joins, as the baseline. %\textsc{Mate} uses the XASH index to filter non-joinable rows. 
\noindent \textbf{Union Search.} 
We select Starmie~\cite{DBLP:journals/pvldb/FanWLZM23}, the most recent and the most effective approach, as the baseline for our union discovery plan. 
% It trains a contrastive model to discover the unionable tables. In addition to column similarities, 
% Starmie leverages column-table associations to learn the semantics of the columns in the context of their tables.
\noindent \textbf{Correlation search.} 
We use the sketch-based correlation approximation approach~\cite{DBLP:conf/icde/SantosBMF22} as the baseline for our correlation discovery plan. This baseline is the most efficient correlation search approach. 
% It creates a join/correlation index that allows simultaneous join and correlation lookup. 

\subsection{Complex Discovery Tasks} \label{sub:complex_tasks}
\begin{table*}[t]
\footnotesize
\caption{Experimental results on complex discovery tasks. B-NO does not use the plan optimizer.}
\centering
\begin{tabular}{l|rrr||rrr||rrr||rrr}
 & \multicolumn{3}{c||}{\textit{\textbf{{With Negative Examples}}}}& \multicolumn{3}{c||}{\textit{\textbf{{Data Imputation}}}}& \multicolumn{3}{c||}{\textit{\textbf{{Feature Discovery}}}}& \multicolumn{3}{c}{\textit{\textbf{{Multi-Objective Discovery}}}}
 \\
% \hline
\textbf{} & \textbf{\system}& \textbf{B-NO} & \textbf{Baseline} & \textbf{\system} & \textbf{B-NO}& \textbf{Baseline}& \textbf{\system}& \textbf{B-NO} & \textbf{Baseline} &\textbf{\system}& \textbf{B-NO} & \textbf{Baseline} \\
\hline
% Table Quality & \textbf{49\%} & 36\% & \textbf{72\%} & 40\% & 100\% & 100\% \\
Runtime & \textbf{14.2} &111.9& 30.5& \textbf{0.19} &0.26& 0.67 & \textbf{9.0} &16.3& 22.8 & \textbf{5.5} &\textbf{5.5}& 47.0  \\
LOC & 5 &5& 72& 5 &5& 51 & 7 &7& 49 & 8 &8& 135 \\
\# of Systems  & 1 &1& 1& 1 &1& 2 & 1 &1& 2 & 1 &1& 3 \\
\# of Indexes & Single& Single & Multi & Single& Single & Multi & Single& Single & Multi &Single & Single& Multi \\
%User Involvement & Low & High & Low & Medium & Low & Medium & Low& Very High \\
% Synthetic & 45\% & 45\% & 69\% & 69\% \\
\hline
\end{tabular}
\label{tab:complex_tasks}
\vspace{-.4cm}
\end{table*}
We first evaluate the capabilities of \system in facilitating the creation of several complex pipelines, namely, \textit{data discovery with negative examples}, \textit{data imputation}, \textit{multicollinearity-aware feature discovery}, and \textit{multi-objective data discovery}. 
While these pipelines may not cover every possible discovery plan that one can implement using \system, they represent commonly discussed discovery applications in the state-of-the-art~\cite{muthukrishnan2016lasso, abedjan2015dataxformer, DBLP:conf/sigmod/YakoutGCC12, DBLP:conf/sigmod/ZhangI20, DBLP:journals/pvldb/ZhangI19}.
%rjm Infogather~\cite{DBLP:conf/sigmod/YakoutGCC12}, DataXFormer~\cite{abedjan2015dataxformer}, and Gen-T~\cite{DBLP:conf/icde/FanSM24}.
%rhn papers discuss data discovery with a given set of either positive or negative examples. Data imputation is widely discussed in data transformation papers~\cite{DBLP:journals/pvldb/ZhuHC17, abedjan2015dataxformer, DBLP:conf/btw/ozmenEA21, singh2012learning}. Although, Multicollinearity is commonly discussed in feature engineering papers~\cite{muthukrishnan2016lasso}, Gen-T proposes this idea in data discovery, where the discovered tables 
% while exuding high similarity to the query, they 
%rjm must not highly correlate with each other.
 %Aurum~\cite{DBLP:conf/icde/FernandezAKYMS18} and Juneau~\cite{DBLP:conf/sigmod/ZhangI20, DBLP:journals/pvldb/ZhangI19} also discuss bringing multiple objectives in one system, but in a more limited way than \system and without optimization.

We implement each task once with \system and once with a federation of available state-of-the-art systems. We further compare the runtime with and without optimization in \system. \textit{\textbf{B-No}} represents \system without optimization.
The results are shown in Table~\ref{tab:complex_tasks}. 
% Then, we evaluate the cost of deploying each approach. 
\subsubsection{Metrics}
We compare the runtime, lines of code (LOC), the number of systems involved in the implementation, and the required number of index structures. 

\subsubsection{Data Discovery with Negative Examples}
In this task, we want to discover tables with disjoint information by specifying negative examples that must be absent in the discovered tables. 

\noindent\textbf{Baseline implementation.} Here, we use \textsc{Mate} and application code. \textsc{Mate} filters irrelevant tables based on a set of examples. The remaining tables are then validated in application code row-by-row to drop tables with negative examples.

\noindent\textbf{\system implementation.} We use two MC seekers and a \textit{Difference} combiner. The first seeker discovers tables based on the query table and the second seeker finds tables comprising the negative examples, which are then filtered by the combiner. 

\noindent\textbf{Results.}
We use the SANTOS benchmark with $365$ queries. We leverage the ground truth unionable tables that do not completely overlap with the query tables and use a random set of non-overlapping rows as negative examples resulting in $1k$ negative examples on average. 
\system is on average twice as fast as the baseline. The row-by-row validation phase is the main bottleneck of the baseline and B-No.
\system rewrites the SQL queries using tables containing negative examples, filtering the irrelevant tables in the database and benefiting from the DBMS-level optimization. 
Although the baseline has only one data discovery system, the user needs to develop a validation phase on top of \textsc{Mate}. 
This leads to over $14$ times more LOC for baseline with $72$ LOC compared to only $5$ LOC required to define the plan in \system.

\subsubsection{Example-Based Data Imputation}
%\textit{Data imputation} is a well-known problem in data science where missing values are automatically generated. 
A common strategy to find missing values is to utilize functional dependencies~\cite{abedjan2015dataxformer}. This requires discovering tables that contain the non-missing row values as well as rows that contain the incomplete rows. 
%To find the tables that most probably contain the missing value, one needs to obtain the intersection of these tables. 

\noindent\textbf{Baseline implementation.} As an ad-hoc solution, one can leverage \textsc{Mate} to first identify tables that contain complete rows, and Josie to discover partial rows. The intersection of the fetched tables is used to infer the missing values.

\noindent\textbf{\system implementation.} Using \system, we simulate the same process with the data imputation sub-plan in Fig.~\ref{fig:pipeline_example}. 

\noindent\textbf{Results.} We select over $1,000$ random column pairs from Gittables, consider the first $5$ rows as the complete examples, and delete the values of the second column for the remaining rows.
As shown in Table~\ref{tab:complex_tasks}, 
\system is over $3.5X$ faster than the baseline. The optimization step re-writes the MC operator to limit the search space to only tables that contain the query values.
The two approaches also differ in terms of LOC.
\system only requires $5$ lines of code
% to implement the discovery plan 
compared to $51$ that are necessary to align the libraries of \textsc{Mate} and Josie, each written in different language. Also, the baseline requires two sets of incompatible indexes to enable data imputation.

\subsubsection{Multicollinearity-Aware Feature Discovery}
%Becktepe et al.~\cite{DBLP:conf/sigmod/BecktepeEKA23} demonstrate a data discovery pipeline 
% concerning a downstream ML task. The goal is to 
%which aims at finding multi-column joinable tables containing correlating features with a target column.
% To ensure that the obtained features provide new information that does not exist in the input dataset, we extend this task by incorporating the multicollinearity requirement.
% , which enforces that the obtained features do not correlate with the given input features.
This is the task of obtaining new features to enrich a dataset for a prediction task. The new features should correlate with the input target but not with each other.
The QCR approach~\cite{DBLP:conf/icde/SantosBMF22} discovers correlating features with a single column. To avoid multicollinearity of the features we need to further expand this. 

\noindent\textbf{Baseline Implementation.} 
To avoid multicollinearity of features, we apply multiple rounds of the QCR approach, once to discover correlating features with the target and then for every feature in the input dataset. Each time we filter the tables that were results of the previous iteration.
In addition, we use \textsc{Mate} to ensure joinability with the join column sets. The final output is the intersection of these two sets of tables.

\noindent\textbf{\system Implementation.} Using \system, one can accordingly use multiple correlation seekers and one MC seeker. The filtering between the correlation seekers is enacted by using a \textit{Difference} combiner after each multicollinearity check.

\noindent\textbf{Results.} We chose $6$ datasets from the SANTOS lake with two joinable string columns and two numerical columns each, one feature and one target.
As shown in Table~\ref{tab:complex_tasks}, \system is $2.5X$ faster than the baseline and $1.8X$ faster than B-NO. The baseline also needs $49$ lines of application code for alignment and filtering, compared to just $7$ lines with \system.

\subsubsection{Multi-Objective Data Discovery}
Our last use case is a subset of the complex plan described in Section~\ref{sub:composing_discovery_task}. 
We exclude data imputation, as we have evaluated it before.

\noindent\textbf{Baseline Implementation.} This pipeline requires combination of three systems Josie~\cite{DBLP:conf/sigmod/ZhuDNM19}, Starmie~\cite{DBLP:journals/pvldb/FanWLZM23}, and QCR~\cite{DBLP:conf/icde/SantosBMF22}.
This needs building three sets of indexes, each requiring a different process from building massive posting lists to training a contrastive model.
Then, one has to compile these approaches, each written in different language, Josie in \texttt{GO}, Starmie in \texttt{Python}, and QCR baseline in \texttt{JAVA}. 
%In addition, each index should be stored differently.
Starmie vectors are stored as a file, QCR performs the best if stored in a column store and Josie is tightly coupled with %data types in
PostgreSQL. 

\noindent\textbf{\system Implementation.} 
The implementation of the multi-objective discovery is shown in Listing~\ref{code:augmentation_by_example}.
Note that we exclude data imputation (Lines~\ref{codeline_augmentation_addmc}-\ref{codeline_augmentation_combiner}).
%Note that all sub-plans are connected via a \textit{Union} combiner, which will not benefit from optimization.

\noindent\textbf{Results.} \system requires fewer lines of code and a single index to implement this complex discovery plan, demonstrating its ability to facilitate the development of complex and on-demand pipelines without either data- or code-level complexities.
\system and the baseline yield $5.5$ and $47$ seconds runtime, respectively. \system is more efficient because it pushes the operators down to the database and prevents time-consuming data loading between the database and memory. As expected, the runtime for \system and B-NO are equal for this plan, because the sub-plans are connected via a \textit{Union} combiner.

% \subsubsection{Summary for complex tasks}
% The experiments demonstrate \system's ability to unburden users in designing complex pipelines, benefiting them with fewer LOC, lower deployment costs, and seamless integration of incompatible systems. 
\subsection{The Optimizer}\label{sub:optimizer_evaluation}
We measure the runtime improvement and the accuracy achieved through rule- and ML-based optimization. 

\subsubsection{Metrics}
We compare the execution time for three approaches: without the optimizer (executing seekers in a random order), with the optimizer, and with an oracle optimizer that consistently identifies the faster seeker. The reported runtimes for \system also include the optimization overhead.
Finally, the \textit{Accuracy} shows the percentage of cases in which the optimizer correctly selects the faster seeker.
The ground truth is obtained through an exhaustive search by executing all combination of operator ranking and labeling the fastest order of seekers as the positive case and the remaining cases as negatives.

The results are summarized in Table~\ref{tab:ml_optimizer}. Each experiment involves $100$ random plans of two seekers connected with an \textit{Intersection} combiner. Gittables serves as the target lake as well as the source of random inputs for the seekers. \textit{Seeker} represents the type of seekers within the plans. 
\begin{table}
\vspace{-.2cm}
\caption{Experiment on optimizer effectiveness.}
\vspace{-.2cm}
\centering
\scriptsize
\begin{tabular}{l|rrr|rr|rr}
& \multicolumn{3}{c|}{\textit{\textbf{{Runtime (Seconds)}}}}& \multicolumn{2}{c|}{\textit{\textbf{{Runtime Gain}}}}& \multicolumn{2}{c}{\textit{\textbf{{Accuracy}}}}
\\
\hline
\textbf{Seeker} & \textbf{Rand} & \textbf{\system} & \textbf{Ideal} & \textbf{\system} & \textbf{Ideal}& \textbf{\system} & \textbf{Ideal} \\
\hline
Mixed & 5.1 & 2.0 & 1.2 & 61.1\% & 75.9\% & 84.4\%& 100\% \\
SC & 1.3 & 1.0 & 0.9 & 21.5\% & 26.9\% & 99.8\% & 100\% \\
MC & 17.2 & 4.5 & 3.1 & 73.7\% & 82.2\%  & 70.3\% & 100\% \\
C & 5.2 & 1.3 & 1.1 & 74.8\% & 77.8\%  & 89.5\% & 100\% \\

% Mixed & 5.1 $\pm$ 11.9366 & 2.0 $\pm$ 5.9563 & 1.2 $\pm$ 2.7027 & 61.1\% $\pm$ 43.35\% & 75.9\% $\pm$ 33.69\% & 84.4\%& 100\% \\
% SC & 1.3 $\pm$ 0.4924 & 1.0 $\pm$ 0.2997 & 0.9 $\pm$ 0.2844 & 21.5\% $\pm$ 16.41\% & 26.9\% $\pm$ 11.13\% & 99.8\% & 100\% \\
% MC & 17.2 $\pm$ 25.4728 & 4.5 $\pm$  24.1566 & 3.1 $\pm$ 7.3494 & 73.7\% $\pm$ 55.83\% & 82.2\% $\pm$ 36.80\% & 70.3\% & 100\% \\
% C & 5.2 $\pm$ 9.6697 & 1.3 $\pm$ 1.5562 & 1.1 $\pm$ 1.3106 & 74.8\% $\pm$ 40.40\% & 77.8\% $\pm$ 33.0\%  & 89.5\% & 100\% \\

\hline
\end{tabular}
\label{tab:ml_optimizer}
\vspace{-.5cm}
\end{table}

\subsubsection{Rule-Based Optimizer}
The first row in Table~\ref{tab:ml_optimizer} shows the effect of the rule-based optimizer, because the seeker types are different. The results show that the rules optimally re-order the seekers in $84.4\%$ of the cases. This leads to a runtime gain of $61.1\%$ over the baseline with random order.

\subsubsection{ML Optimizer}
The last three rows in Table~\ref{tab:ml_optimizer} highlight the efficacy of the ML optimizer.
% , when the seeker types are identical. 
%The results show that 
% \system achieves significantly better runtime compared to the random baseline. O
On average, \system achieves a runtime improvement of $21.5\%$, $73.7\%$, and $74.8\%$ for SC, MC, and correlation seekers, respectively. 
The gap between the ideal runtime and \system's runtime is bigger for the MC seeker. 
%This is mainly because the runtime for MC also depends on the number of columns in $Q$ and the application-level code. 
While our model accounts for the number of columns, it is not possible to predict the exact number of obtained candidate joinable rows and the number of validations required in the application code.

% The reasons \system does not achieve the ideal runtime gains are two fold. 
% First, unlike the ideal approach, the results for \system include the runtime overhead for the optimizer.
% This overhead of on average $0.04$ seconds includes measuring the input size, accessing the frequency dictionary for each value, and predicting the runtime using the ML model.
% Second, the ML model fails to select the faster seeker when the predicted runtime for seekers is in close proximity.

\subsubsection{Assessing the Reliability of the Optimizer}
Similar to database optimizers, our optimizer is not $100\%$ accurate and can lead to an incorrect order of operators~\cite{DBLP:conf/ssdbm/NgWMN99}. 
To assess the reliability of our optimizer, we conducted a statistical hypothesis test on its observed accuracy and runtime improvement.

The null hypothesis ($H_0$) is that our optimizer's superior accuracy is due to chance, while the alternative hypothesis ($H_1$) suggests that the optimizer is significantly better than random. 
Using a z-test, we compare our optimizer's average success rate of $86\%$ ($\hat{p}$) across $4000$ ($n$) queries against the random success rate of $50\%$ under the null hypothesis (($p_0$)). Z-test is calculated as: $\frac{\hat{p} - p_0}{\sqrt{\frac{p_0(1 - p_0)}{n}}} \approx 45.6$.
% $\frac{\hat{p} - p_0}{\sqrt{\frac{p_0(1 - p_0)}{n}}} = \frac{0.86 - 0.50}{\sqrt{\frac{0.50(1 - 0.50)}{4000}}} \approx 45.6$.
Obtaining the z-value, we can calculate: p-value $\ = 2 \times (1 - \Phi(|z|)) \approx 0$, where $\Phi$ is the cumulative distribution function of the normal distribution. This let us to reject the null hypothesis, concluding that our optimizer's superiority is statistically significant.
%Similarly, the z-score for runtime gain shows that the $58\%$ average runtime gain of \system is statistically significant (p-value $\approx 0$) and not due to chance.
% compared to the $0\%$ of the random approach

\subsection{Single-Column Join Search}
We compare our join discovery operator to Josie. 
% As Josie is designed to work with PostgreSQL, we have created an additional column store version for comparison. 
We run the experiments on the WDC, Canada-US-UK Open Data, and the Gittables data lake. 
We follow the same approach as the reference paper~\cite{DBLP:conf/sigmod/ZhuDNM19} to generate query workloads, resulting in $3{,}000$ query columns per data lake, $1{,}000$ per query size.
% We only evaluate the runtime because both approaches are exact join discovery solutions.

\subsubsection{Metrics}
We evaluate both approaches based on runtime, precision@k and recall@k.
WDC and open data lakes are not labeled, therefore, they are only used for the runtime experiments. We use LakeBench benchmark~\cite{DBLP:journals/pvldb/DengCCYCYSWLCJZJZWYWT24} to measure the effectiveness of the approaches.

\subsubsection{Results}
Fig.~\ref{fig:system_vs_josie_runtime} illustrates the results. 
To indicate the storage format, we append \textit{(Column)} (for column-store) and \textit{(Row)} (for PostgreSQL) to the approach names.
% To indicate whether the data lake is stored in the column-store or PostgreSQL, we append \textit{(Column)} and \textit{(Row)} to the name of the approaches, respectively.
The x-axis shows the maximum query size of each query batch, which varies due to different table sizes in the lakes.
% The x-axis in each figure shows the maximum query size of each query batch. Query sizes differ for different data lakes due to their varying table sizes. 
\begin{figure*}[t!]
\vspace{-.3cm}
\begin{tikzpicture}[scale=1.0]
  \begin{groupplot}[xtick=data, group style={group size=3 by 1 , horizontal sep=1.0cm}, height=3.8cm, width=5.5cm, every node near coord/.append ,nodes near coords, point meta=explicit symbolic, ymode = log, log origin=infty, ytick={.01, .1, 1, 10, 100, 1000}, enlarge x limits=0.3]%style={yshift=-0.21cm}

        \nextgroupplot [xtick=data, ybar, symbolic x coords={100, 1k, 10k}, group style={group size=1 by 1 , horizontal sep=.0cm}, every node near coord/.append style={yshift=-0.21cm}, point meta=explicit symbolic, log origin=infty, xticklabel style={font=\footnotesize, rotate=0}, xmajorgrids=true,
    	ymajorgrids=true,grid style=dashed, ylabel={Runtime (seconds)}, x label style={at={(0.5, 1.45)}, anchor=north},
    legend style={legend columns=1,at={(3.9, 1.0)},anchor=north,font=\footnotesize}, xlabel={WDC}, bar width=5pt, ymax = 100, ymin = 0]

% OUR WEBTABLE
% \addplot[blue,fill] coordinates { (10, 0.11581284136566498) (100, 0.14113265649236814) (1000, 0.13591020900385858) };
% \addplot[red,fill] coordinates { (10, 0.02400161270631352) (100, 0.03149499242955988) (1000, 0.02806555185066517) };

\addplot[black,fill] coordinates { (100, 1.9939572629928588) (1k, 2.720255861520767) (10k, 3.59261602973938) };
\addplot[red,fill] coordinates { (100, 0.5952526502609253) (1k, 0.5770588889122009) (10k, 0.854078616142273) };
\addplot[blue,fill] coordinates { (100, 0.1930274052619934) (1k, 0.8726131045818328) (10k, 0.9619960463047028) };
\addplot[orange,fill] coordinates { (100, 5.416967183) (1k, 13.92157328) (10k, 45.08200098) };

\addlegendentry{\system (Row)};
\addlegendentry{Josie (Row)};
\addlegendentry{\system (Column)};
\addlegendentry{Josie (Column)};

    \nextgroupplot [xtick=data, ybar, symbolic x coords={1k, 10k, 100k}, group style={group size=1 by 1 , horizontal sep=.0cm}, every node near coord/.append style={yshift=-0.21cm}, point meta=explicit symbolic, log origin=infty, xticklabel style={font=\footnotesize, rotate=0}, xmajorgrids=true,
    	ymajorgrids=true,grid style=dashed, xlabel={Canada-US-UK}, x label style={at={(0.5, 1.45)}, anchor=north}, bar width=5pt, ymax = 300, ymin = 0.01]

\addplot[black,fill] coordinates { (1k, 0.17081698918342592) (10k, 0.6926211907863616) (100k, 1.4829489605426789) };
\addplot[red,fill] coordinates { (1k, 0.11269223618507385) (10k, 0.44923046016693113) (100k, 2.0529155502319334) };
\addplot[blue,fill] coordinates { (1k, 1.0582309322357177) (10k, 1.2180875525474548) (100k, 2.518535149097443) };
\addplot[orange,fill] coordinates { (1k, 10.226297048146357) (10k, 63.06736845424376) (100k, 170) };

     \nextgroupplot [xtick=data, ybar, symbolic x coords={10, 100, 1000}, group style={group size=1 by 1 , horizontal sep=.0cm}, every node near coord/.append style={yshift=-0.21cm}, point meta=explicit symbolic, log origin=infty, xticklabel style={font=\footnotesize, rotate=0}, xmajorgrids=true,
    	ymajorgrids=true,grid style=dashed, xlabel={Gittables}, x label style={at={(0.5, 1.45)}, anchor=north}, bar width=5pt, ymax = 3, ymin = 0]

\addplot[black,fill] coordinates { (10, 0.019227034568786622) (100, 0.21131408023834228) (1000, 0.18645612215995788) };
\addplot[red,fill] coordinates { (10, 0.014545241395035782) (100, 0.046582162117642975) (1000, 0.04922100572555975) };
\addplot[blue,fill] coordinates { (10, 0.0913811782458881) (100, 0.09781238187252163) (1000, 1.2479657452439912) };
\addplot[orange,fill] coordinates { (10, 0.2642175431251526) (100, 1.2656173288559598) (1000, 2.2077562856624056) };

  \end{groupplot}
\end{tikzpicture}
 \vspace{-.7cm}
\caption{Average runtime comparison between \system and JOSIE for different query sizes ($k$ = $10$).}
\label{fig:system_vs_josie_runtime}
% \vspace{-.4cm}
\end{figure*}
Fig.~\ref{fig:system_vs_josie_runtime} shows that the column-store version of \system consistently outperforms Josie. 
This is because Josie lacks some optimizations that are tightly coupled with specific data types on PostgreSQL.
In the PostgreSQL experiments, except for very large (over $100k$) queries on the Canada-US-UK lake, Josie outperforms \system. Still, the runtime difference is on average below a second.

We also evaluated our approach on a webtable benchmark along with its ground truth provided by the LakeBench paper~\cite{DBLP:journals/pvldb/DengCCYCYSWLCJZJZWYWT24}. Fig.~\ref{fig:lakebench_experiments_all}.a depicts the runtime comparison between \system, Josie, and DeepJoin~\cite{DBLP:journals/pvldb/Dong0NEO23}. 
The results show that DeepJoin achieves the best runtime thanks to its efficient HNSW index. 
Regarding the effectiveness, \system and Josie achieve the same results as their outputs are identical. 
As DeepJoin is a semantic join discovery approach, it achieves higher precision and recall compared to \system and Josie that are designed to discover equi-joins.
% The results reveal that the precision@K of both approaches is over $80\%$ for $k=5$ and $60\%$ for $k=20$ and recall@K is $10\%$ for $k=5$ and $55\%$ for $k=20$. 
% Low recall is due to the large number of joinable tables and a limited $k$ at the same time. Fig.~\ref{fig:lakebench_experiments_all}.b shows the result of this experiment.
% \answerRthree{The ground truth was published in the LakeBench paper~\cite{DBLP:journals/pvldb/DengCCYCYSWLCJZJZWYWT24}.}

% \input{charts_and_tables/new_experiments/lakebench_runtime}
% \input{charts_and_tables/new_experiments/lakebench_precision_recall}
\begin{figure}[t!]
\vspace{-.2cm}
\begin{tikzpicture}[scale=1.0]
  \begin{groupplot}[xtick=data, group style={group size=3 by 1 , horizontal sep=1.5cm}, height=4cm, every node near coord/.append ,nodes near coords, point meta=explicit symbolic,  enlarge x limits=0.2]%style={yshift=-0.21cm}

        \nextgroupplot [xtick=data, width=3.5cm, ybar, symbolic x coords={Webtable Large}, group style={group size=1 by 1 , horizontal sep=.0cm}, every node near coord/.append style={yshift=-0.21cm}, point meta=explicit symbolic, log origin=infty, xticklabel style={font=\footnotesize, rotate=0}, xmajorgrids=true, xlabel={a) Runtime},
    	ymajorgrids=true, grid style=dashed, ylabel={Seconds}, x label style={at={(0.5, 0)}, anchor=north},
    legend style={legend columns=1,at={(0.25, 1.3)},anchor=north,font=\footnotesize}, bar width=10pt, ymax = 5, ymin = 0, ytick={0, 1, 2, 3, 4, 5}, ylabel near ticks]

     \addplot[red, fill] coordinates {(Webtable Large, 3.29)}; \addlegendentry{Josie};
     \addplot[olive, fill] coordinates {(Webtable Large, 0.49)};\addlegendentry{DeepJoin};
     \addplot[blue, fill] coordinates {(Webtable Large, 0.85)};\addlegendentry{\system};

    \nextgroupplot [xtick=data, ybar, symbolic x coords={5, 10, 15, 20}, group style={group size=1 by 1 , horizontal sep=.0cm}, every node near coord/.append style={yshift=-0.21cm}, point meta=explicit symbolic, log origin=infty, xticklabel style={font=\footnotesize, rotate=0}, xmajorgrids=true, xlabel={b) Effectiveness},
    	ymajorgrids=true,grid style=dashed, x label style={at={(0.5, 0)}, anchor=north}, bar width=1.7pt, ymax = 1, ymin = 0, legend style={legend columns=3,at={(0.40, 1.35)},anchor=north,font=\footnotesize}, ytick={.0, .2, .4, .6, .8, 1}, width=5.8cm, ylabel={$\%$}, ylabel near ticks]

     \addplot[blue, fill] coordinates {(5, .83) (10, .75) (15, .69) (20, .6)};
     \addlegendentry{P@k \system};
     \addplot[olive, fill] coordinates {(5, .9) (10, .82) (15, .73) (20, .63)};
     \addlegendentry{P@k DeepJoin};
     \addplot[red, fill] coordinates {(5, .83) (10, .75) (15, .69) (20, .6)};
     \addlegendentry{P@k Josie};
     \addplot[cyan, fill] coordinates {(5, .1) (10, .27) (15, .42) (20, .55)};
     \addlegendentry{R@k \system};
     \addplot[teal, fill] coordinates {(5, .1) (10, .29) (15, .45) (20, .59)};
     \addlegendentry{R@k DeepJoin};
     \addplot[purple, fill] coordinates {(5, .1) (10, .27) (15, .42) (20, .55)};
     \addlegendentry{R@k Josie};

  \end{groupplot}
\end{tikzpicture}
 \vspace{-.6cm}
\caption{Lakebench experiments.}
\label{fig:lakebench_experiments_all}
\end{figure}
%Josie's cost-based filtering is less effective on massive queries. Thus, the practical difference in runtime between Josie and \system is negligible. 

%The experiment also shows that the performance of \system can slightly vary with regard to the data layout of the underlying database. \system performs better in the column store database for all experiments for the WDC data lake and for the Gittables lake when the query size equals $100$. In all other cases, the row store performs better because the column store adds overhead for segmentation when the lake exceeds a certain size. In such cases, instead of pipelining the \textit{group by} of the seeker it uses a hash-based version, which is significantly slower. 
%Increasing the size of query columns leads to a longer runtime for all data lakes and approaches in both databases. 

% Finally, Fig.~\ref{fig:system_vs_josie_runtime_k} illustrates that \system leads to almost identical runtime for different $k$ values because it calculates the joinability of all candidate tables and then fetches the top-$k$ results. On the contrary, Josie's filtering strategies strongly depend on the $k$ value. The smaller the $k$ value, the more Josie can benefit. 

% \input{charts_and_tables/new_experiments/join_runtime_k}

\subsection{Multi-Column Join Discovery}
We compare the MC operator in \system to \textsc{Mate}. 
We use the same benchmark as in the original paper~\cite{DBLP:journals/pvldb/EsmailoghliQA22}, namely DWTC and German open data. The benchmark contains $450$ query tables randomly selected from each lake. 

\subsubsection{Metrics}
We compare both approaches based on their runtime, the average number of true positives (TP), false positives (FP), and precision. A TP is a row that is joinable to an input row on the defined composite key. A FP is a row that does not fully
%rjm something that doesn't overlap at all is also a FP
%only partially 
overlap with the join key  on all columns and therefore is not joinable with any input row.
To find the ground truth, we validate every candidate row by comparing the exact values to the corresponding row in the query table.
\subsubsection{Results}
\system is over $2.6X$ and $10X$ faster than \textsc{Mate} on DWTC and German open data, respectively. 
It is faster because the SQL statement in the MC seeker has a higher filtering effect than the baseline.
%To understand this runtime gap, 
% we evaluate the number of intermediate false positive rates. The only difference between the approaches is the SQL statement, where they fetch the candidate rows. W
%we measure the precision of the approaches by evaluating the number of false positive, i.e., non-joinable, and true positive, i.e., joinable, candidate rows that reach the final time-consuming validation phase. 
To understand this runtime gap, Table~\ref{fig:mcjoin_precision} shows the average TP, FP, and precision for both approaches. 
Recall for both approaches is $100\%$ due to bloom filter character. 
\system achieves over $99\%$ precision on both lakes. %This is because the SQL statement used in \system, leverages all of the join key values to discover candidate rows. 
%This way, the number of candidate rows fetched from the database is dramatically reduced. 
In particular, the average number of obtained rows is $23{,}326$ and $2{,}237{,}671$ for \system and \textsc{Mate} respectively for DWTC, and $5{,}746$ and $215{,}780$ for open data. 
%These results show that the SQL statement in \system can obtain up to two orders of magnitude fewer false positives in the first step of the join discovery. 

\begin{table}[]
     \footnotesize
    \centering
    \vspace{-.2cm}
    \caption{Precision comparison of \system and MATE.}
    \vspace{-.2cm}
    \label{fig:mcjoin_precision}
\begin{tabular}{r|rrr|rrr}
%\toprule

\textit{\textbf{Lake}} & \multicolumn{3}{c|}{\textit{\textbf{{\system}}}}&  \multicolumn{3}{c}{\textit{\textbf{{MATE}}}}
 \\

&\multicolumn{1}{c}{\textbf{TP}}&\multicolumn{1}{c}{\textbf{FP}}&\multicolumn{1}{c|}{\textbf{Precision}}&\multicolumn{1}{c}{\textbf{TP}}&\multicolumn{1}{c}{\textbf{FP}}&\multicolumn{1}{c}{\textbf{Precision}}
\\
\toprule
Open Data  & 5,563 &  \textbf{14} &  \textbf{99.7\%} &  5,563 &  3,587 &  61\% \\
DWTC  & 21,228 &  \textbf{3} &  \textbf{99.99\%} &  21,228 &  7,953 &  73\% \\
\toprule
\end{tabular}
\vspace{-0.5cm}
\end{table}

\subsection{Union Search}\label{sub:exp_union}
We compare the union discovery plan of \system to Starmie. %In this experiment, we only measure the performance of \system on both DBMSs because Starmie does not use a DBMS.
The experiment involves four data lakes from the Starmie paper~\cite{DBLP:journals/pvldb/FanWLZM23}: SANTOS, SANTOS Large, TUS, and TUS Large. They contain $50$, $80$, $150$, and $100$ query tables, respectively.

\begin{figure}[t!]
\centering
\vspace{-.2cm}
\begin{tikzpicture}
  \begin{groupplot}[xtick=data, ybar, enlarge x limits=0.2, symbolic x coords={SANTOS, SANTOS Large, TUS, TUS Large}, ymin = 0, group style={group size=1 by 1 , horizontal sep=.0cm}, height=4.2cm, width=8cm, ymax = 200, every node near coord/.append style={yshift=-0.21cm}, point meta=explicit symbolic, xticklabel style={font=\footnotesize}, xmajorgrids=true, ytick={0, 1, 10, 100}, ymode = log, ymin = 0, log origin=infty,
    	ymajorgrids=true,grid style=dashed ]
    \nextgroupplot [ylabel={Runtime (seconds)}, 
    legend style={legend columns=1,at={(0.78, 1.0)},anchor=north,font=\footnotesize}
    , bar width=6pt]
     \addplot[red, fill] coordinates {(SANTOS, 1.8) (SANTOS Large, 19) (TUS, 2.9) (TUS Large, 9.6) }; \addlegendentry{STARMIE};
     \addplot[black, fill] coordinates {(SANTOS, 41) (SANTOS Large, 160) (TUS, 7.7) (TUS Large, 8.1)};\addlegendentry{\system (Row)};
     \addplot[blue, fill] coordinates {(SANTOS, 4.7) (SANTOS Large, 30.2) (TUS, 4.6) (TUS Large, 2.7)};\addlegendentry{\system (Column)};
     % \addplot[orange, fill] coordinates {(SANTOS, .43) (SANTOS Large, 20.8) (TUS, .44) (TUS Large, .72)}; \addlegendentry{\system (DuckDB)};
  \end{groupplot}
\end{tikzpicture}
\caption{Time comparison between \system and STARMIE.}
 \vspace{-.4cm}
\label{fig:union_runtime}
\end{figure}

\subsubsection{Metrics}
In addition to the runtime of the approaches, similar to the previous works~\cite{DBLP:journals/pvldb/FanWLZM23, DBLP:journals/pacmmod/KhatiwadaFSCGMR23, DBLP:journals/pvldb/NargesianZPM18}, we use precision@k, recall, and Mean Average Precision@k (MAP) to measure the quality of the results. For the quality metrics, for each $k$ and data lake, the \textbf{bold} value shows the approach with a higher score. 
We use the ground truth introduced in the original papers~\cite{DBLP:journals/pvldb/FanWLZM23, DBLP:journals/pvldb/NargesianZPM18, DBLP:journals/pacmmod/KhatiwadaFSCGMR23}.

\subsubsection{Results}
Fig.~\ref{fig:union_runtime} shows that Starmie outperforms \system in all cases except TUS Large due to its fast contrastive model and HNSW index to compute the embedding distances in memory. 
% Note that Starmie keeps the data in memory.
\system (Column) is one order of magnitude faster than \system (Row). 
As Starmie considers both syntactic similarities and semantic associations, the final results of the approaches are different. Therefore, we also compare the quality of the results. 
% According to the benchmark provided by the baseline papers, 
Note that SANTOS Large does not contain ground truth labels.
According to Table~\ref{tab:union_quality}, Starmie slightly outperforms \system for $k=10$. This is because some unionable tables with high semantic relatedness significantly lack overlap similarity. For $k=20$, \system and Starmie perform similarly. For higher $k$ values \system achieves better results in all cases. 
%Increasing $k$, the MAP for Starmie decreases from $98\%$ for $k=10$ to $96\%$, $91.5\%$, and $87.5\%$ for $k=20$, $k=50$ and $k=100$, respectively, while \system maintains consistent precision, recall, and MAP across all $k$ values.
Both approaches display low recall for small $k$ values on the TUS and TUS Large lakes.
This is because the ground truth for these benchmarks has a large number of unionable tables.% i.e., over $163$ and $269$ tables on average. %while the possible results are bound by the significantly smaller $k$. 
For $k=10$, the ideal recall is $6\%$ and $4\%$ for TUS and TUS Large respectively. 
%Although increasing $k$ to $50$ and $100$ increases the recall in both approaches because they can select more tables, \system reaches higher recall faster than Starmie. 
Overall, the results show that \system's native union operator is competitive with state-of-the-art.

\begin{table*}[]
     \footnotesize
    \centering
    \vspace{-.2cm}
    \caption{Quality comparison between \system and STARMIE.}
     \vspace{-.2cm}
    \label{tab:union_quality}
\begin{tabular}{r|rrr|rrr||rrr|rrr}
%\toprule

 & \multicolumn{6}{c||}{\textit{\textbf{{k=10}}}}& \multicolumn{6}{c}{\textit{\textbf{{k=20}}}}
 \\

\textit{\textbf{Lake}} & \multicolumn{3}{c|}{\textit{\textbf{{\system}}}}&  \multicolumn{3}{c||}{\textit{\textbf{{STARMIE}}}}& \multicolumn{3}{c|}{\textit{\textbf{{\system}}}}&  \multicolumn{3}{c}{\textit{\textbf{{STARMIE}}}}
 \\

&\multicolumn{1}{c}{\textbf{P@k}}&\multicolumn{1}{c}{\textbf{Recall}}&\multicolumn{1}{c|}{\textbf{MAP}}&\multicolumn{1}{c}{\textbf{P@k}}&\multicolumn{1}{c}{\textbf{Recall}}&\multicolumn{1}{c||}{\textbf{MAP}}
&\multicolumn{1}{c}{\textbf{P@k}}&\multicolumn{1}{c}{\textbf{Recall}}&\multicolumn{1}{c|}{\textbf{MAP}}&\multicolumn{1}{c}{\textbf{P@k}}&\multicolumn{1}{c}{\textbf{Recall}}&\multicolumn{1}{c}{\textbf{MAP}}\\
\toprule
SANTOS  & 92\% &  70\% &  94\% &  \textbf{97\%} &  \textbf{73\%} &  \textbf{99\%} &  \textbf{100\%} &  89\% &  94\% &  97\% &  \textbf{95\%} &  \textbf{98\%} \\
TUS  & 94\% &  05\% &  94\% &  \textbf{95\%} &  \textbf{06\%} &  \textbf{98\%} &  \textbf{95\%} &  10\% &  94\% &  92\% &  10\% &  \textbf{96\%} \\
TUS Large & \textbf{94\%} &  04\% &  94\% &  93\% &  04\% &  \textbf{97\%} &  \textbf{94\%} &  07\% &  94\% &  90\% &  07\% &  94\% \\
\toprule
 & \multicolumn{6}{c||}{\textit{\textbf{{k=50}}}}& \multicolumn{6}{c}{\textit{\textbf{{k=100}}}}
 \\
TUS  & \textbf{96\%} & \textbf{24\%} & \textbf{95\%} & 90\% & 23\% & 93\% & \textbf{93\%} & \textbf{43\%} & \textbf{92\%} & 81\% & 38\% & 90\%\\
TUS Large & \textbf{92\%} & \textbf{17\%} & \textbf{93\%} & 85\% & 15\% & 90\% & \textbf{92\%} & \textbf{32\%} & \textbf{92\%} & 77\% & 26\% & 85\%\\
 
\toprule
\end{tabular}
\vspace{-.3cm}
\end{table*}

\subsection{Correlation Discovery}\label{sub:correlation_evaluation}
We compare the correlation operator of \system with the QCR baseline~\cite{DBLP:conf/icde/SantosBMF22} on the NYC lake. As our approach avoids the mechanics for random sampling during runtime and follows a convenience sampling approach, we also test a version where the indexed rows are apriori shuffled. We call it \system (rand).
The first benchmark from the QCR paper, i.e., NYC (Cat.), comprises queries with one categorical join key and one numerical target column. For the second benchmark, i.e., NYC (All), we allow for the join key to be selected from any column type, not restricted to only categorical columns. 
This is because the join keys are not restricted to categorical columns, e.g., all the keys in the TPC-H benchmark are numerical.

\subsubsection{Metrics}
We measure the runtime, effectiveness (precision@k and recall@k), and ablate the sampling method.
To find the ground truth, we calculate the exact correlation between the query target and each candidate from the data lake and identify the top-$k$ correlating columns.

\subsubsection{Results}
%We generate two query benchmarks comprised of randomly selected pairs of columns containing a join key and target column.

Table~\ref{tab:correlation_precision_recall} shows the P@$k$ and recall for \system, \system (rand), and the baseline for $k=10$ and $h=256$.
% as the hash size. 
On the NYC (All) benchmark, \system outperforms the baseline by $18\%$ and $17\%$ for P@$k$ and R@$k$, respectively. 
This performance gap is because the baseline cannot discover the correlating columns when the join key is numeric. 
On NYC (Cat.), the baseline achieves slightly higher P@$k$ and recall. This is due to their differences in hashing and sketching. Further, random sampling (\system (random)) outperforms the vanilla \system, which uses convenience sampling based on \texttt{LIMIT h} sorted by RowId. As data sorted by RowId follows an unknown order, consecutive values might be duplicates making this sampling non-representative.

The runtime of systems is very similar: On NYC (Cat.) the average runtime in seconds is $0.67$ and $0.86$ and on NYC (All) $0.56$ and $1.33$ for the baseline and \system, respectively. 
%This illustrates that \system's index structure leads to similar efficiency as the state-of-the-art approach. 
%Both the runtime and quality performance of \system slightly decrease for NYC (All) compared to NYC (Cat.). This is because of the higher number of candidate tables in the former benchmark, which causes longer runtime and lower precision and recall due to the increased number of correlation calculations.

% In summary, our sketching strategy simplifies the index drastically.
% to serve the objective of this paper in building a unified data discovery system. 
% This leads to a $32\%$ smaller index size compared to the baseline. Nevertheless, our index 
% addresses more general tasks. It 
In summary, the correlation seeker of \system is competitive while being flexible with regard to the type of the join column and also allowing sketch sizes to be specified on-demand.
% , facilitating the end user in balancing runtime and quality trade-offs for different correlation discovery tasks.

% \begin{table}
% \caption{Experiment on Correlation-based discovery.}
% \centering
% \begin{tabular}{l|rr||rr}
%  & \multicolumn{2}{c||}{\textit{\textbf{{\system}}}}& \multicolumn{2}{c}{\textit{\textbf{{Baseline~\cite{DBLP:conf/icde/SantosBMF22}}}}}
%  \\
% \hline
% \textbf{Lake} & \textbf{P@10} & \textbf{R@10} & \textbf{P@10} & \textbf{R@10} \\
% \hline
% NYC (All) & \textbf{45\%$\pm$34\%} & \textbf{42\%$\pm$33\%} & 26\%$\pm$34\%& 26\%$\pm$ 33\%\\
% NYC (Cat.) & 55\%$\pm33\%$ & 57\%$\pm$34\% & \textbf{64\%$\pm$32\%} & \textbf{61\%$\pm$31\%} \\
% % Synthetic & 45\% & 45\% & 69\% & 69\% \\
% \hline
% \end{tabular}
% \label{tab:correlation_precision_recall}
% \end{table}

\begin{table}
\caption{Experiment on Correlation-based discovery.}
\vspace{-.2cm}
\centering
\footnotesize
\begin{tabular}{l|rr||rr||rr}
& \multicolumn{2}{c||}{\textit{\textbf{{\system}}}}& \multicolumn{2}{c||}{\textit{\textbf{{\system (rand)}}}}& \multicolumn{2}{c}{\textit{\textbf{{Baseline~\cite{DBLP:conf/icde/SantosBMF22}}}}}
\\
\hline
\textbf{Lake} & \textbf{P@10} & \textbf{R@10} & \textbf{P@10} & \textbf{R@10} & \textbf{P@10} & \textbf{R@10} \\
\hline
NYC (All) & 42\% & 40\% & \textbf{44\%}&\textbf{42\%}& 24\%& 23\%\\
NYC (Cat.) & 55\% & 57\% &60\%&58\%&\textbf{64\%} & \textbf{61\%} \\
% Synthetic & 45\% & 45\% & 69\% & 69\% \\
\hline
\end{tabular}
\label{tab:correlation_precision_recall}
\vspace{-.3cm}
\end{table}

\subsection{Index Size in \system}
Table~\ref{tab:index_storage_complexity} compares the storage required for \system's index and the combination of DataXFomer, Josie, \textsc{Mate}, Starmie, and QCR indexes. 
The index for WDC and Canada-US-UK lakes excludes the QCR index as they lack table identifiers. 
On average, \system requires $57\%$ less storage compared to the combination of the state-of-the-art indexes.
The index construction time of \system ranged between $2$ to $80$ hours for Gittables and DWTC, respectively, which is dominated by the inverted index creation. 
% Note that index generation is a singular offline process.
\begin{table}[]
    \footnotesize
    \centering
    \caption{The storage required for \system and S.O.T.A.} %a combination of indexes introduced in the state-of-the-art (S.O.T.A.).}
    \vspace{-.2cm}
    \label{tab:index_storage_complexity}
\begin{tabular}{r|l|l}
\toprule
\textit{\textbf{Data lake}} & \textit{\textbf{\system}} & \textit{\textbf{Combination of S.O.T.A indexes}} \\ \toprule
DWTC                        &     \textbf{998.0 GB} & 3.6 TB \\
Gittables                   &     \textbf{397.3 GB} & 511.0 GB \\
German Open Data            &      \textbf{79.7 GB} & 575.7 GB \\
WDC                         &     \textbf{151.8 GB} & 223.7 GB (W/O QCR)\\
Canada, US, and UK &    \textbf{114.7 GB} & 230.1 GB (W/O QCR)\\
TUS                         &       \textbf{7.9 GB} &   8.4 GB \\
TUS Large                   &      \textbf{12.2 GB} &  13.1 GB \\
SANTOS                      &       \textbf{5.5 GB} &   6.0 GB \\
SANTOS Large                &     \textbf{135.0 GB} & 144.2 GB \\
NYC open data               &     \textbf{709.2 GB} & 763.0 GB
\end{tabular}
%\vspace{-.4cm}
\end{table}

\subsection{User Study}\label{sub:userstudy}

\begin{table}
\vspace{-.2cm}
\caption{Statistics obtained from the conducted user study.}
\vspace{-.2cm}
\centering
\resizebox{\columnwidth}{!}{
\tiny
\begin{tabular}{l|c|c|c}
%& \multicolumn{3}{c}{\textit{\textbf{{Results}}}}\\
\hline
& \textbf{Research} & \textbf{Industry} & \textbf{All} \\
\hline
Number of \textbf{participants} & 9 & 9 & 18 \\
\hline
\multicolumn{4}{l}{\cellcolor{gray!20}\textit{\textbf{{Question 1. How often do you find data within a single search?}}}}\\
% \textbf{I find data within a single search (1 (rarely) - 5 (often))} & 2.1& 2.55 & 2.33 \\
\textbf{Rarely (0\%) - Often (100\%)} & 27.5\%& 38.8\% & 33.3\% \\
\hline
\multicolumn{4}{l}{\cellcolor{gray!20}\textit{\textbf{{Question 2. Is a single discovered table sufficient as the output of the discovery task?}}}}\\
\textbf{Yes} $|$ \textbf{No} & 11\% $|$ \textbf{89\%}& 00\% $|$ \textbf{100\%} & 06\% $|$ \textbf{74\%} \\
\hline
\multicolumn{4}{l}{\cellcolor{gray!20}\textit{\textbf{{Question 3. What are your most frequent data discovery tasks?}}}}\\
\textbf{Discovery for rows}  & 33\%& \textbf{67\%}& \textbf{50\%} \\
\textbf{Correlation discovery}  & \textbf{44\%}& 56\%& \textbf{50\%}\\
\textbf{Join discovery}  & \textbf{44\%}& 33\%& 39\%\\
\textbf{Keyword search}  & \textbf{44\%}& 33\%& 39\%\\
\textbf{multi-column join discovery}  & 33\%& 22\%& 28\%\\
\hline
\multicolumn{4}{l}{\cellcolor{gray!20}\textit{\textbf{{Question 4. How do you solve data discovery tasks?}}}}\\
\textbf{With custom scripts}& \textbf{100\%}& \textbf{56\%}& \textbf{78\%}\\
\textbf{Writing SQL queries}& 44\%& \textbf{56\%}& 50\% \\
\textbf{Asking people}& 33\% & \textbf{56\%} & 44\%\\
\textbf{Using open source tools} & 56\%& 33\% & 44\% \\
\textbf{Using commercial tools} & 22\%& 22\% & 22\% \\
\hline
\multicolumn{4}{l}{\cellcolor{gray!20}\textit{\textbf{{Question 5. What programming language do you prefer?}}}}\\
% \textbf{Python} $|$ \textbf{Java} $|$ \textbf{SQL} $|$ \textbf{C++} &\textbf{100\%} $|$ 78\% $|$ 78\% $|$ 56\%& \textbf{89\%} $|$ \textbf{89\%} $|$ 78\% $|$ 78\%& \textbf{94\%} $|$ 83\% $|$ 78\% $|$ 67\%\\
\textbf{Python} &\textbf{100\%} & \textbf{89\%} & \textbf{94\%} \\
\textbf{Java} $|$ \textbf{SQL} $|$ \textbf{C++} & 78\% $|$ 78\% $|$ 56\%& \textbf{89\%} $|$ 78\% $|$ 78\%&83\% $|$ 78\% $|$ 67\%\\
\hline
\multicolumn{4}{l}{\cellcolor{gray!20}\textit{\textbf{{Question 6. Where do you store your data lake?}}}}\\
\textbf{DBMS} $|$ \textbf{File systems} $|$ \textbf{Both} & 33\% $|$ \textbf{44\%} $|$  22\% & 44\% $|$ 00\% $|$ \textbf{56\%} & \textbf{39\%}  $|$ 22\% $|$\textbf{39\%} \\
% \textbf{DBMS} & 33\% & 44\% & \textbf{39\%} \\
% \textbf{File systems} & \textbf{44\%}& 00\% & 22\% \\
% \textbf{Both} &  22\% & \textbf{56\%} & \textbf{39\%} \\
\hline
\multicolumn{4}{l}{\cellcolor{gray!20}\textit{\textbf{{Question 7. Would you use DBMS if indexing and optimizations are provided?}}}}\\
\textbf{YES} $|$ \textbf{NO} & \textbf{100\%} $|$ 0\%& \textbf{100\%} $|$ 0\% & \textbf{100\%} $|$ 0\% \\
\hline
\multicolumn{4}{l}{\cellcolor{gray!20}\textit{\textbf{{Question 8. Which API do you prefer for \underline{simple} tasks?}}}}\\
\textbf{\system} $|$ \textbf{Python} $|$ \textbf{SQL} & 34\% $|$ 22\% $|$ \textbf{44\%} & \textbf{56\%} $|$ 11\%  $|$ 34\% & \textbf{44\%} $|$ 17\% $|$ 39\% \\
% \textbf{\system}& 34\% & \textbf{56\%}  & \textbf{44\%} \\
% \textbf{Python} & 22\% &  11\%  &  17\% \\
% \textbf{SQL} & \textbf{44\%} & 34\% & 39\% \\
\hline
\multicolumn{4}{l}{\cellcolor{gray!20}\textit{\textbf{{Question 9. Which API do you prefer for \underline{complex} tasks?}}}}\\
\textbf{\system} $|$ \textbf{Python} &  \textbf{89\%} $|$ 11\%   &  \textbf{89\%} $|$ 11\%  & \textbf{89\%} $|$ 11\% \\
% \textbf{\system} &  \textbf{89\%}&  \textbf{89\%} & \textbf{89\%} \\
% \textbf{Python} & 11\%   &  11\%  & 11\% \\
\hline
\end{tabular}
\label{tab:survey_results}
}
\vspace{-.4cm}
\end{table}
To assess the usefulness of our system and the validity of our premises we conducted a survey and obtained results from $18$ expert participants ($40\%$ response rate) covering sectors, such as healthcare, banking, technology, and education. The detailed results can be found on our GitHub repository\footnote{\url{https://github.com/LUH-DBS/Blend/blob/main/UserStudy.md}}.
%Through this survey, we aim to understand preferences, needs, and challenges in data discovery. 
% We quiz participants on a variety of aspects.

A concise representation of the results of the user responses is depicted in Table~\ref{tab:survey_results}.
% Here, we summarize the survey results.
The survey reveals that our expert participants rarely find desired datasets within a single search (question $1$).
Rather, their discovery needs are often met by composition of tables (question $2$).
%With regard to the prevalence of discovery types, participants commonly selected complex tasks. 
The two most common tasks were discovering tables \textit{containing a set of rows} and \textit{containing a correlating column to a target}. 
These tasks are more prevalent in industry (question $3$). 
$78\%$ of the participants solve such problems via custom scripts from scratch, with $94.4\%$ using Python (questions $4$ and $5$). 
Moreover, $78\%$ indicated that their data lakes are managed at least partly, via DBMS; $39\%$ reported to only use DBMS (question $6$). 
Industrial participants showed a higher preference for storing the data lake in a DBMS compared to academics.
This is because research often prioritizes expeditious results, while in industry the emphasis is on stable and maintainable solutions.
All respondents unanimously expressed that they would use a DBMS if data discovery capabilities such as, indexes and optimizations are provided (question $7$). 
The answers to the previous questions underpin the motivation for our approach.

We also surveyed users on their preference for simple and complex task implementations. 
The simple task involved querying tables with specific keywords, while 
the complex task aimed at finding joinable tables on two columns that also have correlating columns to a target. 
For both tasks, we provided implementations via \system and alternatives. 
The simple task was presented in \system's API, Python, and SQL, while the complex task could only be implemented using \system's API and Python.
For the simple task, the majority ($44.4\%$) favor \system followed by SQL ($38.9\%$). 
For the complex task, this gap is larger ($89\%$ vs. $11\%$ in favor of \system), as the Python implementation becomes significantly complex. 
Participants justify their choice through \system's more flexible, concise, easy-to-understand, and robust API (questions~$8$ and~$9$).
%The questions and the concrete tasks can be found within the full survey report on our project GitHub.
% \subsection{Summary}
% \system enables users to construct custom data discovery plans through a single API, eliminating the need for multiple solutions. It provides a user-friendly interface to integrate various discovery solutions into a unified framework. Experiments show that \system achieves similar to or better performance than individually optimized counterparts, utilizing the database $45\%$ more than baselines on average. This efficiency is because \system moves $83.5\%$ of discovery computations into the database, reducing data loading time by $54.5\%$ for certain operations compared to alternatives such as \textsc{Mate}. For more details, we refer to the results on our GitHub.

\section{Related Work}\label{sec:related}
Our system \system relates to several lines of research.

\noindent\textbf{Data discovery in data lakes.} Several large companies have reported about internal platforms and engines to retrieve structured information from their in-house data lakes. 
LinkedIn published
WhereHows~\cite{bworld}, Microsoft introduced DLN~\cite{DBLP:journals/pvldb/BharadwajGBG21}, the Apache foundation offers Atlas~\cite{atlas}, and
Google presented Goods~\cite{DBLP:conf/sigmod/HalevyKNOPRW16} and Google Dataset Search~\cite{DBLP:conf/www/BrickleyBN19}. These systems record the life-cycle of datasets through metadata and lineage, where discovery operators are generally underexplored.

Fernandez et al.~\cite{DBLP:conf/icde/FernandezAKYMS18} presented Aurum, a data discovery system utilizing an enterprise knowledge graph to connect data lake tables based on joinability. 
The system is capable of answering keyword queries. It also allows for further exploration of the joinable tables based on pre-calculated join similarities in the data lake. Moreover, Aurum provides a domain-specific language to facilitate browsing of related tables. 
Because of the precaluclation, Aurum lacks the capability for ``table as query'' search. \system provides more low-level operators beyond single-column join discovery and keyword search that can be declaratively composed to a plan and optimized.

%Zhang et al.~\cite{DBLP:journals/pvldb/ZhangI19, DBLP:conf/sigmod/ZhangI20} introduce Juneau, a discovery system designed for Jupyter notebooks. It uses pipelines, provenance data, and intermediate results to find relevant data for the user's current workflow. %Juneau generates a dependency graph based on the provenance information of previously used notebooks and identifies the most similar datasets or preprocessing techniques that a user may require in developing their pipeline.
%Unlike Juneau, \system is a general-purpose system. In theory, Juneau could run on top of \system.

\noindent\textbf{Individual Discovery Techniques.}
In addition to the aforementioned approaches, several studies introduce and utilize index structures for discovery tasks, e.g., union search~\cite{DBLP:conf/www/BayardoMS07, DBLP:journals/pvldb/NargesianZPM18, DBLP:conf/icde/BogatuFP020, DBLP:journals/pvldb/FanWLZM23, DBLP:conf/sigmod/SarmaFGHLWXY12, DBLP:conf/sigmod/ZhangI20}, join search~\cite{DBLP:conf/sigmod/ZhangI20, DBLP:conf/sigmod/SarmaFGHLWXY12, DBLP:conf/sigmod/ZhuDNM19, DBLP:journals/tods/XiaoWLYW11, DBLP:conf/icde/XiaoWLS09, DBLP:conf/edbt/VenetisSR12, DBLP:conf/icde/FernandezMNM19, DBLP:journals/pvldb/ZhuNPM16, DBLP:journals/pvldb/CasteloRSBCF21, DBLP:journals/pvldb/SuriIRR21, DBLP:journals/pvldb/EsmailoghliQA22, DBLP:conf/sigmod/Sarawagi04, DBLP:conf/icde/FernandezAKYMS18, DBLP:journals/pvldb/CafarellaHK09, DBLP:conf/www/BayardoMS07}, keyword search~\cite{DBLP:conf/www/BrickleyBN19, DBLP:conf/cikm/ZhangSF21, DBLP:journals/pvldb/CafarellaHK09, DBLP:conf/icde/FernandezAKYMS18}, correlation search~\cite{DBLP:conf/edbt/EsmailoghliQA21, DBLP:conf/sigmod/BecktepeEKA23, DBLP:conf/icde/SantosBMF22, DBLP:conf/sigmod/SantosBCMF21, DBLP:journals/pvldb/ChepurkoMZFKK20}, and augmentation by example~\cite{abedjan2015dataxformer, DBLP:conf/sigmod/YakoutGCC12, DBLP:journals/pvldb/KhatiwadaSGM22, DBLP:conf/cikm/AmsterdamerC21,DBLP:conf/icde/GalhotraGF23}.
Among these techniques, some are closely related to our techniques.
%D3L~\cite{DBLP:conf/icde/BogatuFP020}, Table Union %Search~\cite{DBLP:journals/pvldb/NargesianZPM18}, and 
Starmie~\cite{DBLP:journals/pvldb/FanWLZM23} discovers unionable tables using an ensemble of indexes.
%D3L leverages five distinct structures, each facilitating specific similarity computations.
%Similarly, Table Union Search proposes three LSH indexes, capturing content, semantic, and metadata similarities. 
It leverages a contrastive model and a fast HNSW index to find the most similar column embeddings.
Josie~\cite{DBLP:conf/sigmod/ZhuDNM19} leverages two indexes for efficient join discovery using a set of data-dependent pruning techniques.
COCOA~\cite{DBLP:conf/edbt/EsmailoghliQA21} uses a combination of inverted index and \textit{Order index} to find non-linear correlating columns from data lake in linear time. %Infogather~\cite{DBLP:conf/sigmod/YakoutGCC12, DBLP:conf/sigmod/ZhangC13} leverages five indexes to augment data by examples or metadata.
MATE~\cite{DBLP:journals/pvldb/EsmailoghliQA22} is the only approach that addresses the problem of multi-column join discovery. It introduces a hash-based inverted index and a filtering strategy that facilitates discovery of relevant tables based on a composite join key.
Each of these approaches address a single task. In \system, we combine indexes of the prevalent tasks. 

\noindent\textbf{Indexes for semantic data discovery.} 
To facilitate semantic data discovery~\cite{DBLP:conf/icde/DongT0O21, DBLP:journals/pvldb/CasteloRSBCF21, DBLP:journals/pvldb/SuriIRR21, DBLP:journals/pvldb/BharadwajGBG21, DBLP:journals/fcsc/YuLDF16, DBLP:conf/sigmod/LiHDL15, DBLP:journals/sigmod/WandeltDGMMPSTWWL14, DBLP:journals/pvldb/FanWLZM23}, indexes for high dimensional embeddings are proposed. 
HNSW~\cite{DBLP:conf/cikm/MaTL23, DBLP:journals/pvldb/FanWLZM23} and IVF-PQ~\cite{DBLP:conf/sigir/ONeillD23} are indexes to accelerate vector search in finding similar elements embedded as high-dimensional vectors. Although these structures enable efficient 
%rjm discovery of approximate joins, 
nearest neighbor search on vectors, they are not easy to integrate within our holistic token-based index. Further, their approximate nature make their inclusion into our optimizer non-trivial as reordering might change the result set.

\noindent\textbf{Optimizing concurrent queries.} Multi-query optimizations (MQO)s~\cite{DBLP:conf/bigdataconf/TuEXC22, DBLP:journals/tods/Sellis88, DBLP:conf/edbt/HongRKGD09, DBLP:journals/dke/ShimSN94} aim to create a single global query to avoid duplicate sub-queries. Our seekers contain complex predicates compared to simple point queries, making global query creation impractical due to rare overlaps in queries. Our optimizer uses query rewrites that uses intermediate results to limit the search space for subsequent queries, regardless of overlaps in their logical plans.

\noindent\textbf{DAGs and ETL.}
Our optimization strategies assume a topological order of operator groups. Existing work on ETL optimization also deals with 
DAGs~\cite{DBLP:journals/concurrency/GuCQZG24, DBLP:conf/caise/DongenDM08, DBLP:conf/iccbr/MinorTB07} that connect input schemata to output schemata via specific activities~\cite{SimitsisVS05,SimitsisVS05TKDE}. While we borrow ideas from ETL and DB optimization, as our discovery pipelines also can be modeled as DAGs, we are dealing with an entirely different optimization goal. The underlying data in our scenario is unknown to the user. The optimization potential is guided by the user query and its join and unionability with the data that resides inside the lake. Further our table discovery operations go beyond relational operators that are studied for optimization in ETL pipelines.
\section{Conclusion and Future Work} \label{sec:conclusion}
We propose a unified system for customized data discovery tasks that allows for declarative task definition.
To this end, we introduce atomic operators to serve as building blocks for higher-level tasks. 
Furthermore, we present a minimal yet general index that facilitates efficient execution of operators. 
To decrease data serialization, we push-down the operators into database by implementing them in SQL. 
We also propose a task-level optimization for efficient query execution.
Our system does not cover semantic and fuzzy relationships. It would be interesting to extend our system to enable the execution and optimization of these operators. 
This can include incorporation of high-dimensional embeddings into our index structure. The use of in-DB embeddings would also enable efficient vector indexing using methods like HNSW or IVFFlat.
% This extension will be valuable in the existence of erroneous data lakes.

% \section*{Acknowledgment}

% % The preferred spelling of the word ``acknowledgment'' in America is without 
% % an ``e'' after the ``g''. Avoid the stilted expression ``one of us (R. B. 
% % G.) thanks $\ldots$''. Instead, try ``R. B. G. thanks$\ldots$''. Put sponsor 
% % acknowledgments in the unnumbered footnote on the first page.
% We thank Matthias Böhm for his insightful feedback. 
% This work was supported by he German Research Foundation (DFG) under grant agreement ... and the German Ministry for Education and Research as BIFOLD — “Berlin Institute for the Foundations of Learning and Data” (...).

% \begin{acks} 
% This project has been supported by the German Research Foundation (DFG) under grant agreement 387872445 and the German Ministry for Education and Research as BIFOLD — “Berlin Institute for the Foundations of Learning and Data” (01IS18025A and 01IS18037A).
% \end{acks}

\balance

% \small
%\bibliographystyle{abbrv}
% \bibliographystyle{ACM-Reference-Format}
\bibliographystyle{abbrv}
\bibliography{references}

\end{document}